%% file: sigproc-sp.tex
\documentclass{acm_proc_article-sp}

\usepackage{qtree}
\usepackage{verbatim} 
\usepackage{framed}
\usepackage{alltt}
\usepackage{pepa}
\usepackage{stackrel}
\usepackage{amsmath}
\usepackage{amsfonts}
\usepackage{graphicx}

\usepackage{amsthm} 
\usepackage{semantic}
\usepackage{float}
\usepackage[margin=1cm,scriptsize,sc,centerlast]{caption}
\DeclareCaptionType{copyrightbox}
\usepackage[font=small,skip=2pt]{subcaption}
\usepackage[cmtip,all]{xy}
\usepackage{color}
\usepackage{hyperref}

\floatstyle{boxed}
\newfloat{program}{htp}{lop}
\floatname{program}{Program}
\floatstyle{plain}
\newfloat{model}{thp}{lop}
\floatname{model}{Model}

\newtheoremstyle{mystyle}
{-5pt} 
{-10pt} 
{} 
{} 
{\bfseries} 
{.} 
{.5em} 
{} 

\theoremstyle{mystyle}
\newtheorem{mydef}{Definition}
\newtheorem{condition}{Condition}

\newcommand{\mysync}[1]{\stackbin[#1]{}{\rhd\!\!\!\lhd}}

\newcommand{\group}[1]{\mathcal{G}(\mo{#1})}
\newcommand{\smgroup}[1]{\mathcal{G}_{s}(\mo{#1})}
\newcommand{\lagroup}[1]{\mathcal{G}_{l}(\mo{#1})}
\newcommand{\mo}[1]{\mathbb{#1}}

\newcommand{\interface}[2]{\mathcal{I}(\mo{#1},#2)}
\newcommand{\subsid}[3]{\mathcal{J}(#1,#2,#3)}
\newcommand{\einterface}[2]{\mathcal{I_\mathcal{A}}(#1,#2)}

\newcommand{\coop}[3]{sync(#1,#2,#3)}

\newcommand{\actions}[1]{ \stackrel{\rightarrow}{\mathcal{A}^{*}}\!\!(#1) }
\newcommand{\actionsI}[1]{ \stackrel{\rightarrow}{\mathcal{A}^{*}_{i}}\!\!(#1) }
\newcommand{\actionsLarge}[1]{\stackrel{\rightarrow}{\mathcal{A}^{*}_{l}}\!(#1)}
\newcommand{\actionsLargeModel}[1]{\stackrel{\rightarrow}{\mathcal{A}^{*}_{l}}\!(\mo{#1})}
\newcommand{\actionsSmallModel}[1]{\stackrel{\rightarrow}{\mathcal{A}^{*}_{s}}\!(\mo{#1})}
\newcommand{\actionsSmallLargeModel}[1]{\stackrel{\rightarrow}{\mathcal{A}^{*}_{sl}}\!(\mo{#1})}

\newcommand{\actionsLargeModelNormal}[1]{\stackrel{\rightarrow}{\mathcal{A}^{*}_{l}}\!({#1})}
\newcommand{\actionsSmallModelNormal}[1]{\stackrel{\rightarrow}{\mathcal{A}^{*}_{s}}\!({#1})}
\newcommand{\actionsSmallLargeModelNormal}[1]{\stackrel{\rightarrow}{\mathcal{A}^{*}_{sl}}\!({#1})}

\newcommand{\actionsLargeGroup}[1]{\stackrel{\rightarrow}{\mathcal{A}^{*}_{l}}\!\!(#1)}
\newcommand{\actionsSmallGroup}[1]{\stackrel{\rightarrow}{\mathcal{A}^{*}_{s}}\!(#1)}
\newcommand{\actionsSmallLargeGroup}[1]{\stackrel{\rightarrow}{\mathcal{A}^{*}_{sl}}\!(#1)}

\newcommand{\verylongsquiggly}{\xymatrix@C=80pt{{}\ar@{~>}[r]&{}}}
\newcommand{\longsquiggly}{\xymatrix@C=30pt{{}\ar@{~>}[r]&{}}}
\newcommand{\shortsquiggly}{\xymatrix@C=15pt{{}\ar@{~>}[r]&{}}}

\newcommand{\transition}[3]{#1 \stackrel{(#3)}{\longrightarrow}#2}
\newcommand{\longtransition}[3]{#1 \stackrel{(#3)}{\xrightarrow{\hspace*{1.5cm}}}#2}

\newcommand{\proba}[1]{\mathbb{P}_{\langle #1 \rangle}(t)}
\newcommand{\probNoT}[1]{\mathbb{P}_{\langle #1 \rangle}}
\newcommand{\sumArray}[2]{\begin{array}{c}
#1 , \\
#2
\end{array}}

\newtheorem{lemma}{Lemma}[section]

\makeatletter

\newcommand{\Rmnum}[1]{\expandafter\@slowromancap\romannumeral #1@}
\makeatother

\begin{document}

\title{An Aggregation Technique For Large-Scale PEPA Models With Non-Uniform Populations}

\numberofauthors{1} 
\author{
\alignauthor Alireza Pourranjbar , Jane Hillston \\
\affaddr{LFCS, School of Informatics, University of Edinburgh} \\
\email{a.pourranjbar@sms.ed.ac.uk , jane.hillston@ed.ac.uk}
} 

%
\date{}

\maketitle

\begin{abstract}
%
Performance analysis based on modelling consists of two major steps: model construction and model analysis.
Formal modelling techniques significantly aid model construction but can exacerbate model analysis.
In particular, here we consider the analysis of \emph{large-scale} systems which consist of one or more
entities replicated many times to form large populations.
The replication of entities in such models can cause their state spaces
to grow exponentially to the extent that their exact stochastic analysis becomes computationally expensive or even infeasible.

In this paper, we propose a new \emph{approximate aggregation} algorithm for a class of large-scale PEPA models.
For a given model, the method quickly checks if it satisfies a \emph{syntactic} condition, indicating that the model may be solved
approximately with high accuracy.
If so, an aggregated CTMC is generated directly from the model description.
This CTMC can be used for efficient derivation of an approximate marginal probability distribution over some of the model's populations.
In the context of a large-scale client-server system, we demonstrate the usefulness of our method.
\end{abstract}




\input{./Introduction/new-introduction}

\input{./Preliminaries/new-preliminaries}

\input{./AggregationCondition/new-aggregationCondition}

\input{./Algorithm/new-algorithm}

\input{./CalculatingMarginals/new-calculatingMarginals}

\input{./CheckingTheAccuracy/checkingTheAccuracy}

\input{./RelatedWork/new-relatedWork}
\input{./Conclusion/new-conclusion}

\bibliographystyle{abbrv}
\bibliography{myref}  
%
%

\appendix
\section{Rules For the Rates of Passive Activities}
These rules are used when calculating the apparent rates of passive activities. 
\begin{eqnarray*}
\label{eqn:passiveRateRules}
\nonumber1.&& (\omega_1 \top + \omega_2 \top)=(\omega_1 + \omega_2) \top\ \ \ \ \ \ \ \forall \ \  \omega_1,\omega_2 \in \mathbb{N}  \\
2.&& \omega_1 \times (\omega_2 \top) = (\omega_1 \times \omega_2) \top  \\
\nonumber3.&& \top  =  1 \times \top  \\
\nonumber4.&& \min(\omega_1\times\top,\omega_2\times\top) = min(\omega_1,\omega2)\times\top
\end{eqnarray*}

\section{Conversion of the C-K equations}
Consider Model $\mo{M}$ which satisfies Condition~1. Let $X$ be the random process captured by $\mo{M}$\rq{}s original CTMC and $\mathbb{P}_{\langle X=S_i\rangle}(t)$ denote the probability that at time $t$, $X$ is in state $S_i$. In the model's C-K equations, one equation is constructed for each state $X$ experiences: 
\begin{multline}
\!\!\!\!\!\frac{\proba{X = S_i}}{d\ t}  =  
\!\!\!\!\!\!\!\!\!\!\!\!\!\!\!\!\!\!\!\!\!\!\!\!\!\!\!\!\!
\!\!\!\!\sum_{ \ \ \ \ \ \ \ \ \ \ \ \ \ \ \ \ \ \ \longtransition{S_j}{S_i}{\alpha,r_{\langle\alpha,S_j,S_i\rangle}},\ S_j\in Y_{\mo{M}}}^{}
\!\!\!\!\!\!\!\!\!\!\!\!
r_{\langle\alpha,S_j,S_i\rangle}\proba{X=S_j}
\, 
\\[-2pt]
\ \ \ \ \ \ \ \ \ \ \ \ \ \ \ \ \ - \!\!\!\!\!\!\!
\!\!\!\!\!\!\!\!\!\!\!\!\!\!\!\!\!\!\!\!\!\!\!\!\!\!\!
\sum_{\ \ \ \ \ \ \ \ \ \ \ \ \ \ \longtransition{S_i}{S_j}{\alpha,r_{\langle\alpha,S_i,S_j\rangle}}\ ,\ S_j \in Y_\mo{M}}^{} \mkern-18mu \!\!\!\!\! 
\!\!\!\!\!\!\!\!\!\!\!\!\!\!\!\!
r_{\langle\alpha,S_i,S_j\rangle}\proba{X=S_i}
\end{multline}
The probability of being in a sub-chain $Y_i$ at time $t$ is equal to the sum of the probabilities of being in any of the states in $Y_i$:  $\proba{X \subset Y_i} = \sum_{S_i \in Y_i}^{}\ \proba{X=S_i}$.
To study these probability evolutions, we form one equation for each sub-chain $Y_i$. $\proba{X\subset Y_i}$ denotes the probability that at time $t$ the system is in any of the states in $Y_i$.
\begin{eqnarray}
\label{complexCKEquation}
\frac{d\  \proba{X\subset Y_i}}{d\ t} \!= \! \frac{d\ \sum_{S_i \in Y_i}^{}\!\proba{X=S_i} }{d\ t}\! =\!\!\! \sum_{S_i \in Y_i}^{}\!\!\!\frac{d\ \proba{X=S_i}}{d\ t}
\label{asdf}
\end{eqnarray}
Using Eq. (4) in (5) we derive: 
\begin{multline}
\label{eqn:bothComponents}
\!\!\!\!\!\!
\sum_{S_i \in Y_i}^{}\!\!\frac{d\ \proba{X=S_i}}{d\ t} \! = \!
\!\!\!
\sum_{S_i\ \in\ Y_i}^{} \!\Big( 
- 
\!\!\!\!\!\!\!\!\!\!\!\!\!\!\!\!\!\!\!\!\!\!\!\!\!\!\!\!\!\!\!\!\!\!\!\!\!\!\!\!\!\!\!\!\!\!\!\!\!\!\!\!\!\!\!\!\!\!\!
\sum_{
\ \ \ \ \ \ \ \ \ \ \ \ \ \ \ \ \ \ \ \ \ \ \ \ \ \ \ \ \ 
\sumArray{\longtransition{S_i}{S_j}{\alpha,r_{\langle\alpha,S_i,S_j\rangle}}\ ,\ S_j\in Y_j\ }{\alpha \in (\actionsSmallLargeModel{M}\  \cup \actionsSmallModel{M})}}^{}
\mkern-18mu\mkern-18mu\mkern-155mu
 \proba{X=S_i} \times  r_{\langle\alpha,S_i,S_j\rangle} \\
 + 
\!\!\!\!\!\!\!\!\!\!\!\!\!\!\!\!\!\!\!\!\!\!\!\!\!\!\!\!\!\!\!\!\!\!\!\!\!\!\!\!\!\!\!\!\!\!\!\!\!\!\!\!\!\!\!\!\!\!\!
\sum_{
\ \ \ \ \ \ \ \ \ \ \ \ \ \ \ \ \ \ \ \ \ \ \ \ \ \ \ \ \ 
\sumArray{\longtransition{S_k}{S_i}{\alpha,r_{\langle\alpha,S_k,S_i\rangle}}\ ,\ S_k \in Y_k\ }{\alpha\ \in (\actionsSmallLargeModel{M}\ \cup \actionsSmallModel{M})}} 
\mkern-18mu\mkern-18mu\mkern-155mu
\proba{X=S_k} \times  r_{\langle\alpha,S_k,S_i\rangle} \Big)
\end{multline} 
Considering the first term of (6) and Lemma (\ref{observation}): 
\begin{eqnarray}
\label{eqn:firstComponent}
&& \sum_{S_i \in Y_i}{} \Big(
\!\!\!\!\!\!\!\!\!\!\!\!\!\!\!\!\!\!\!\!\!\!\!\!\!\!\!\!\!\!\!\!\!\!\!\!\!\!\!\!\!\!\!\!\!\!\!\!\!\!\!\!\!\!\!\!\!\!\!
\sum_{
\ \ \ \ \ \ \ \ \ \ \ \ \ \ \ \ \ \ \ \ \ \ \ \ \ \ \ \ \ 
\sumArray{\longtransition{S_i}{S_j}{\alpha,r_{\langle\alpha,S_i,S_j\rangle}}\ ,\ S_j \in Y_j\ }{\alpha \in (\actionsSmallLargeModel{M}\  \cup \actionsSmallModel{M})}}^{}
\mkern-18mu\mkern-18mu\mkern-155mu
 \proba{X=S_i} \times  r_{\langle\alpha,S_i,S_j\rangle}\ 
\big) \\ 
\nonumber 
&= &\sum_{S_i\in Y_i}^{} \ \big(
\!\!\!\!\!\!\!\!\!\!\!\!\!\!\!\!\!\!\!\!\!\!\!\!\!\!\!\!\!\!\!\!\!\!\!\!\!\!\!\!\!\!\!\!\!\!\!\!\!\!\!\!\!\!\!\!\!\!\!
\sum_{
\ \ \ \ \ \ \ \ \ \ \ \ \ \ \ \ \ \ \ \ \ \ \ \ \ \ \ \ \ 
\sumArray{\longtransition{S_i}{S_j}{\alpha,r_{\langle\alpha,S_i,S_j\rangle}}\ ,\   S_j \in Y_j\ }{\alpha \in (\actionsSmallLargeModel{M}\  \cup \actionsSmallModel{M})}}^{}
\mkern-18mu\mkern-18mu\mkern-155mu
 \proba{X=S_i} \times r_{\langle\alpha,Y_i,Y_j\rangle}\  \Big)
 \\
\nonumber & = &  \sum_{S_i\in Y_i}^{} \proba{X=S_i}\ \Big(
\!\!\!\!\!\!\!\!\!\!\!\!\!\!\!\!\!\!\!\!\!\!\!\!\!\!\!\!\!\!\!\!\!\!\!\!\!\!\!\!\!\!\!\!\!\!\!\!\!\!\!\!\!\!\!\!\!\!\!
\sum_{
\ \ \ \ \ \ \ \ \ \ \ \ \ \ \ \ \ \ \ \ \ \ \ \ \ \ \ \ \ 
\sumArray{\longtransition{S_i}{S_j}{\alpha,r_{\langle\alpha,S_i,S_j\rangle}}\ ,\   S_j \in Y_j\ }{\alpha \in (\actionsSmallLargeModel{M}\  \cup \actionsSmallModel{M})}}^{}
\mkern-18mu\mkern-18mu\mkern-155mu
  r_{\langle\alpha,Y_i,Y_j\rangle}\ 
\ \ \ \ \ \ \ \ \ \ \ \ \ \ \ \Big) \\
\nonumber & \approx &\sum_{S_i \in Y_i}^{} \proba{X=S_i} \left(
\sum_{{Y_j \in \mathbb{D}}}^{}\   r_{\langle Y_i,Y_j\rangle}\ 
\right)  \\
\nonumber & = & \left(
\sum_{{Y_j \in Y_{\mo{M}}}}^{}\   r_{\langle Y_i,Y_j\rangle}\ 
\right) \sum_{S_i \in Y_i}^{} \proba{X=S_i} \\
\nonumber & = & \left(
\sum_{{Y_j \in Y_{\mo{M}}}}^{}\   r_{\langle Y_i,Y_j\rangle}\ 
\right) \proba{X \subset Y_i}.
\end{eqnarray} 
The approximation step is based on the assumption that the probability of being in boundary states in nearly zero and therefore, the same probability fluxes as of the non-boundary ones can be associated to them. 

Similarly, for the second term of Eq. (\ref{eqn:bothComponents}):

\begin{eqnarray}
\label{eqn:secondComponent}
&&\sum_{S_i \in Y_i}{} \Big(
\!\!\!\!\!\!\!\!\!\!\!\!\!\!\!\!\!\!\!\!\!\!\!\!\!\!\!\!\!\!\!\!\!\!\!\!\!\!\!\!\!\!\!\!\!\!\!\!\!\!\!\!\!\!\!\!\!\!\!
\sum_{
\ \ \ \ \ \ \ \ \ \ \ \ \ \ \ \ \ \ \ \ \ \ \ \ \ \ \ \ \ 
\sumArray{\longtransition{S_k}{S_i}{\alpha,r_{\langle\alpha,S_k,S_i\rangle}}\ ,\ S_k \in Y_k\ }{\alpha \in (\actionsSmallLargeModel{M}\  \cup \actionsSmallModel{M})}}^{}\ 
\mkern-18mu\mkern-18mu\mkern-155mu
\proba{X=S_k} \times  r_{\langle\alpha,S_k,S_i\rangle}\ 
\Big)    \\
\nonumber & = & \sum_{S_i \in Y_i}\   \Big(
\!\!\!\!\!\!\!\!\!\!\!\!\!\!\!\!\!\!\!\!\!\!\!\!\!\!\!\!\!\!\!\!\!\!\!\!\!\!\!\!\!\!\!\!\!\!\!\!\!\!\!\!\!\!\!\!\!\!\!
\sum_{
\ \ \ \ \ \ \ \ \ \ \ \ \ \ \ \ \ \ \ \ \ \ \ \ \ \ \ \ \ 
\sumArray{\longtransition{S_k}{S_i}{\alpha,r_{\langle\alpha,Y_k,Y_i\rangle}}\ ,\ S_k\in Y_k\ }{\alpha \in (\actionsSmallLargeModel{M}\  \cup \actionsSmallModel{M})}}^{}\ 
\mkern-18mu\mkern-18mu\mkern-155mu
\proba{X=S_k} \times  r_{\langle\alpha,S_k,S_i\rangle}\ 
\Big)  \\
\nonumber& \approx & \sum_{S_i \in Y_i}^{}\ \left(
\sum_{\transition{S_k}{S_i}{.,.}\ ,\ S_k\in Y_k}^{}\ \proba{X=S_k} \times  r_{\langle Y_k,Y_i\rangle}\ 
\right) \ \ \ \ 
\begin{array}{c}
\text{} \\
\text{}
\end{array}
\\
\nonumber& = &
\sum_{S_i \in Y_i}\ \sum_{Y_k \in Y_{\mo{M}}}\ \sum_{\transition{S_k}{S_i}{\cdot,\cdot}\ ,\ S_k \in Y_k}
\Big(
\ \proba{X=S_k} \times  r_{\langle Y_k,Y_i\rangle}\ 
\Big) 
\\
\nonumber& = &\sum_{Y_k \in Y_{\mo{M}}}\ \sum_{\transition{S_k}{S_i}{\cdot,\cdot}\ ,\ S_k \in Y_k}
\Big(
\ \proba{X=S_k} \times  r_{\langle Y_k,Y_i\rangle}\ 
\Big)  \\
\nonumber & \approx &\sum_{Y_k \in Y_{\mo{M}}}\ r_{\langle Y_k,Y_i\rangle} \times \sum_{S_k \in Y_k}\ \proba{X = S_k}  \\ 
\nonumber & = &\sum_{Y_k \in Y_{\mo{M}}}\ r_{\langle Y_k,Y_i\rangle} \times \proba{X \subset Y_k}
\end{eqnarray}
The first approximation step here is done through assuming a negligible probability for being in boundary states. Substituting Eq. \ref{eqn:firstComponent} and \ref{eqn:secondComponent} in Eq. \ref{eqn:bothComponents}, we derive: 
\begin{multline}
\forall\  Y_i \in Y_{\mo{M}},  
\frac{d\ \proba{X \subset Y_i}}{d\ t} = 
\\
  - \sum_{Y_j \in Y_{\mo{M}}}\ \proba{X\subset Y_i} \times r(Y_i,Y_j) \\
  + \sum_{Y_k \in Y_{\mo{M}}} \proba{X\subset Y_k} \times r(Y_k,Y_i)
\end{multline}
Eq (9) captures the probability evolutions at the level of sub-chains.

\balancecolumns

\end{document}

%% file: Introduction/new-introduction.tex
\section{Introduction}

Discrete state modelling is a very expressive style of modelling which has been used in a number of domains, including performance modelling, where it is supported by a number of modelling formalisms such as queueing networks, stochastic Petri nets and stochastic process algebras.  These formalisms ease the task of model construction but can exacerbate the underlying disadvantage of discrete state modelling, the problem of state space explosion.  In this paper we present a new aggregation technique for a class of Markovian-based models expressed in the stochastic process algebra PEPA~\cite{Hillston:1996}.  
This formalism supports compositionality: first the behaviours of individual components are defined and then these are composed to form the system's complete description. The aggregation approach we develop takes advantage of this feature. 

In this paper, we consider a class of \emph{large-scale} PEPA models, i.e.\
models in which there exist one or more components which are instantiated many times to form large populations.
In such models, the interactions between individual components can be seen as interactions between the corresponding populations. In particular we focus on models in which component populations have significantly different sizes --- some components appear in large populations, but other components belong to populations which are relatively small. 
The models in this class reflect many resource-bound computer and communication networks, which typically 
consist of two types of entities, \emph{resources} and \emph{resource users}.  In such systems it is often the case that a rather large population of the users are served by a smaller population of the resources. 


Having constructed the model of a large-scale system, theoretically a wide variety of analysis techniques can be applied for their performance evaluation.
At one extreme, numerical analysis would construct the associated infinitesimal generator matrix and obtain the model's probability distribution across its complete state space. Whilst it provides a full and faithful account of the system behaviour this route often proves to be infeasible due to the size of the state space.
At the other end of the spectrum, analysis based on fluid approximation or mean field techniques can be used to derive the probability distribution's first few moments rather than the full  distribution~\cite{Hayden:2010}. 
These moments can provide the basis for performance estimation and the 
fluid flow analysis has specifically been shown to be useful for models with uniformly large populations~\cite{mirco:thesis}.  
However, applying the method to our sub-class of models  can give rise to misleading results~\cite{dontGoWith}, because the fluid flow moments can be too abstract or crude to reflect the system's full distribution (e.g.\ the distribution may be heavy-tailed or multi-modal). 

Here we present an approach which is particularly tailored for studying resource-bound systems in which large populations of users interact with a limited population of resources.
Our method is based on a state space aggregation and the information it provides about the complete probability distribution for resource populations is more detailed and fine-grained than the fluid flow moments. At the same time, it avoids the construction of the model's potentially very large CTMC\@.
We show that for a large-scale model in this sub-class, the aggregation is possible if the model satisfies a syntactic condition which can be readily checked, and from the aggregated CTMC certain performance measures can be readily derived.

Paper's structure: in Section~\ref{preliminaries}, we introduce the syntax and state space representation used when dealing with large-scale PEPA models. In Section~\ref{condition} we formally introduce the sub-class and illustrate an aggregation condition. In Sections~\ref{sec:aggregationAlgorithm} and \ref{equationAlgorithm} we show the aggregation steps and the way a marginal probability distributions is derived. In Section~\ref{checkingTheAccuracy}, we describe the usefulness of the method in the context of an example. Section~\ref{relatedWork} presents the related work. In Section~\ref{conclusion} we conclude and describe plans for extending the method.

%% file: Preliminaries/new-preliminaries.tex
\section{Preliminaries}
\label{preliminaries}

In this section we give a brief introduction to the PEPA modelling language. 
The interested reader is referred to~\mbox{\cite{Hillston:1996,Hayden:2010}} for more detail.  In PEPA, models are constructed of components which undertake activities.  Each activity $(\alpha, r)$ has a specified activity type, $\alpha$, and a rate, $r$, which is assumed to govern an exponentially distributed delay determining the duration of the activity.  Sequential components are composed concurrently, but may be constrained to share particular activities. In some cases the rate of an activity may be unspecified, or \emph{passive}, denoted $\top$, indicating that a component is willing to participate in a shared activity with no constraint on the rate.   Here we use the\emph{grouped} PEPA syntax for clear specification of large-scale PEPA models~\cite{Hayden:2010}.
\begin{eqnarray*}
S & = & (\alpha,r). S\  | \ S + S \  | \ C_s \ \ \ \ \ \ \ P  =  P \mysync{L} P\  |\  S \\ 
D & = & D \ ||\ D \ | \ P  \ \ \ \ \ \ \ \ \ \ \ \  \ \ \ \ \ \ \ M  =  M \mysync{L} M\  |\  Y\{D\}  
\end{eqnarray*}
Here $S$ represents a \emph{sequential} component (essentially, a state machine).  These components are built up using \emph{prefix} (.): designated first activity, and \emph{choice} (+): preemptive alternative behaviours.  $C_s$ is a constant used to name a component.  $P$, a \emph{model} component, is composed of sequential components ($S \mysync {L} S$) which are constrained to share activities whose type is in the set $L$.  When $L$ is empty these components may proceed independently, denoted by $P~||~Q$. Syntactic sugar $P[n]$ is used to denote the parallel composition of $n$ identical $P$ processes, $P[n] = \underbrace{P~||~P~|| \ldots ~||~P}_{n\ \text{times}}$. $D$ represents the \emph{parallel} composition of instances of a sequential component. These instances constitute a group and a unique label is assigned to each group, denoted by $Y$. $M$ represents the composition of groups. We restrict ourselves to grouped PEPA models in which all groups are strictly \emph{simple}; i.e.\ each group contains the instances of one sequential component only. Here is an example of such a model.

\textbf{Model 1}. This model captures the dynamics of a client-server system.
\begin{eqnarray*}
&& \!\!\!\!\!\!\!\!C_{\textit{think}} \rmdef (\textit{think}, r_{t}).C_{\textit{req}} \ \ \ \ \ \ \ \ C_{\textit{req}} \rmdef (\textit{req}, r_{c}).C_{\textit{think}} 
\end{eqnarray*}
\begin{eqnarray*}
&& \!\!\!\!\!\!\!\!S_{idle} \rmdef (req, r_{s}).S_{log} + (brk, r_{b}).S_{broken}\\
&& \!\!\!\!\!\!\!\!S_{\textit{log}}   \rmdef (\textit{log}, r_l).S_{\textit{idle}} \ \ \ \ \ \ \ \ \ \ \ \ \ S_{\textit{broken}} \rmdef (\textit{fix}, r_{f}).S_{\textit{idle}} \\
&& \!\!\!\!\!\!\!\!CS \rmdef \textit{Servers}~\{\ S_{\textit{idle}}[2]\ \}  \mysync{\{ req\}} \textit{Clients}~\{\ C_{\textit{think}}[2]\ \}
\end{eqnarray*}
{\bf Description:} $C_{\textit{think}}$ and $C_{\textit{req}}$ are the local states of the Client sequential component: clients alternate between thinking and requesting.  Similarly $S_{\textit{idle}}, S_{\textit{log}}$ and $S_{\textit{broken}}$ define the behaviour of the Server sequential component. In state $S_{\textit{idle}}$ a server offers the shared activity $\textit{req}$, followed by logging in state $S_{log}$.  Alternatively the server might suffer a breakdown ($\textit{brk}$) and enter the state $S_{\textit{broken}}$ until it is fixed.

The last line of the model is called the \emph{system equation} and corresponds to $M$ in the grammar. This specifies how the sequential components are instantiated in groups and the cooperation pattern between groups. Here, the model is initialised with two instances of idle servers and two instances of thinking clients. The servers form group $\textit{Servers}$ and the group of the clients is labelled $\textit{Clients}$. The cooperation set specifies that instances in group $\textit{Servers}$ synchronise on $\textit{req}$ activities with instances in group $\textit{Clients}$. Moreover, there is no synchronisation between the instances of servers or clients within their respective groups.

The grouped PEPA structured operational semantics can be used to build the models' underlying transition relations~\cite{Hayden:2010}. The relations can help when generating the model's underlying stochastic process, readily seen to be a CTMC. 
For a grouped PEPA model $\mathbb{M}$, let $\group{\mathbb{M}}$ denote its set of group labels. This can be calculated recursively: 
\begin{eqnarray}
\group{\mathbb{M}} & = & \left\{			\begin{array}{ll}
									{M_1} \cup {M_2} &\ \ \ \  \  \text{if} \ \ \   \mathbb{M} = M_1 \mysync{L} M_2   \\
									Y & \ \ \ \  \  \text{if} \ \ \   \mathbb{M} = Y\{D\}
								\end{array} \right. 
\end{eqnarray}
Let $N^{h}_{\mathbb{M}}$ denote the number of groups in $\mathbb{M}$. 
We define $\mathcal{C}_{\mathbb{M}}$ to be the set of sequential components defined in the model. The number of distinct sequential components in the model is denoted by $N^{c}_{\mathbb{M}}$. Given that all groups in $\mathbb{M}$ are simple, we can construct the function $sc(H): \group{\mathbb{M}} \rightarrow \mathcal{C}_{\mo{M}}$ which relates a group's label to the sequential component whose instances form that group\footnote{We assume that a sequential component is represented by its initial state.}. For example, $sc(\textit{Servers}) = S_{\textit{idle}}$. 

For a $C \in \mathcal{C}_{\mathbb{M}}$, let $ds(C)$ denote the set of local states or local derivatives $C$ visits. We assume that the number of local states the sequential component $C_x$ experiences is denoted by $N_{C_x}$. 
As an example, $ds(S_{\textit{idle}})\!\!=\!\!\{S_{\textit{idle}},S_{\textit{log}},S_{\textit{broken}}\}$.
Accordingly, we define $ds^{*}(H)$, the set of local states each instance in the group labelled $H$ experiences. For $H\!\!\in\!\!\mathcal{G}(\mathbb{M})$: $\!\! ds^{*}(H)\!\!=\!\! ds(sc(H))$. $N_{H_i}$ denoted the cardinality of $ds^{*}(H)$. 
In the system equation of a model, the operator $\mysync{L}$ is used to compose the model's groups and form cooperations. 
The instances of two groups which are composed are restricted to synchronize on action types in the associated cooperation set. 
These groups, however, can be subjected to further compositions and synchronizations with other groups of the model. 
Thus, in a model, the system equation introduces a \emph{hierarchy of cooperation} among its different groups and assigns to each group, the set of action types the instances in that group must synchronize on. To clarify, consider the process 
$( G_1\{\cdot\} \mysync{L^{\prime}} G_2\{\cdot\} ) \mysync{L} ( G_3\{\cdot\} \mysync{L^{\prime\prime}} G_4\{\cdot\} )$. 
Here $G_1$ and $G_2$ cooperate on set $L'$;
similarly, $G_3$ and $G_4$ cooperate on actions in set $L''$. 
Additionally, these group compositions are also combined in cooperation on set $L$. 
The cooperation on $L'$ restricts the instances in groups $G_1$ and $G_2$ and does not affect $G_3$ or $G_4$;  but all instances are restricted by $L$.
The notion of cooperation hierarchy in model $\mo{M}$ can be formally captured by a partial order relation $<^{*}_{\mo{M}}$. For group compositions, $G_1$ and $G_2$, \mbox{${G_1}\!\!<^{*}_{\mo{M}}\!\!{G_2}$}, if and only if $G_1$ is composed when constructing~$G_2$. 
If \mbox{${G_1}\!\!<^{*}_{\mo{M}}\!\!{G_2}$}, all cooperation sets applied to $G_2$ are also enforced on $G_1$. 
The following rules construct $<^{*}_\mo{M}$ using $\mo{M}$'s system equation.
\[ \begin{array}{lr}
1 .\ H\{\cdot\}  <^{*}  H\{\cdot\} & 2.\ X <^{*}  A,\ $ if\  \ \ $A \rmdef X\\
\multicolumn{2}{l}{3.\ H\{\cdot\}  <^{*}  {H_1\{\cdot\} \mysync{L} H_2\{\cdot\}}\ ,$  if\ \ $H=H_1 \vee H=H_2}
\end{array} \]
For a group $H\in\group{{M}}$, $\interface{M}{H}$ denotes the set of action types on which instances in $H$ are required to synchronize. $\interface{M}{H}$ is defined recursively from the system equation, using the subsidiary function $\mathcal{J}$; $\interface{M}{H} = \subsid{\mo{M}}{H}{\emptyset}$. 
\begin{eqnarray*}
\subsid{M}{H}{K} & \! =\!& 	\left \{ 
												\begin{array}{ll}
													K,   \ \ \ \ \  \ \ \ \  \ \ \   \ \ \ \ \text{if}\  M = H\{\cdot\}  & \\
													\subsid{M_1}{H}{K \cup L } \cup \subsid{M_2}{H}{K\cup L }, & \\
													 \ \ \ \ \ \ \ \ \ \ \ \ \ \ \ \ \ \  \ \ \text{if} \ \ \ M = M_1 \mysync{L} M_2. & \\ 
													 \subsid{A}{H}{K},  \  \ \ \ \text{if}\ \ \ M \rmdef A.  & \end{array}
											\right.
\end{eqnarray*}
We also define $\coop{\mo{M}}{H}{\alpha}$, the set of groups in $\mo{M}$ whose instances synchronize on $\alpha$ activities with instances in $H$.
Using $<^{*}_{\mo{M}}$, $\coop{\mo{M}}{H}{\alpha}$ can be found as follows:
\begin{eqnarray*}
\coop{M}{H}{\alpha} &\!\!\!\!\!\! =\!\!\!\!\!\! & 
\left\{		\begin{array}{ll}
															\!\!\!\! \emptyset, \ \ \ \ \text{if}\ M = H\{\cdot\} \\[3pt] 
															\!\!\!\! \coop{M_1}{H}{\alpha}\!\cup\!\{H_i\!\in\!\group{M}\! \mid\! H_i\{\cdot\}\!\! <^{*}_{\mo{M}}\!\!M_2\!\}, \\ 
															\ \ \ \  \text{if}\ M=M_1 \mysync{L} M_2, \alpha\in L, H<^{*}_{\mo{M}} M_1 \\[3pt]
															\!\!\!\! \coop{M_2}{H}{\alpha}\!\cup\!\{H_i\!\in\!\group{M}\!\mid\! H_i\{\cdot\}\!\!<^{*}_{\mo{M}}\!\!M_1\!\}, \\
															\ \ \ \  \text{if}\ M=M_1 \mysync{L} M_2, \alpha\in L, H<^{*}_{M} M_2	\\[3pt]
															\!\!\!\! \coop{M_1}{H}{\alpha}, \\ 
															\ \ \ \ \text{if}\ M=M_1 \mysync{L} M_2, \alpha\not\in L, H<^{*}_{\mo{M}} M_1 \\[3pt] 
															\!\!\!\! \coop{M_2}{H}{\alpha}, \\
															 \ \ \ \ \text{if}\ M=M_1 \mysync{L} M_2, \alpha\not\in L,  H<^{*}_{\mo{M}} M_2
\\[3pt]
														\end{array}\right.
\end{eqnarray*}
\textbf{State Space Representation.}
In the context of large-scale PEPA models, we are often content to abstract from the behaviour of individuals and focus on the evolution of populations~\cite{1595779}. Thus, in order to capture the current state of a group $H$, we do not capture the state of each individual instance but instead, for each local state $C_{x,y}$ in $ds^{*}(H)$, 
with a counter $\xi(H,C_{x,y})$ we record how many instances in $H$ exhibit the behaviour associated with $C_{x,y}$ and form the vector $\boldsymbol{\xi}(H)\!=\!\langle\xi(H,C_{x,y})\mid C_{x,y}\!\in\!ds^{*}(H)\rangle$ to capture the distribution of instances in $H$ across the local states $ds^{*}(H)$.
Having formed $\boldsymbol{\xi}(H)$ for each group, we construct $\boldsymbol{\xi}\!\!=\!\!\langle \boldsymbol{\xi}(H)\mid H \in \group{M}\rangle$, the vector which captures the state of all groups in the model.
The model's \emph{dimension}, $d_{\mo{M}}\!\!=\!\!\sum_{H\in\group{M}}^{}N_{H}$, is defined to be the number of state variables appearing in its state vector.

Each counter $\xi(H,C_{x,y})$ can be regarded as a random variable and therefore, the state vector is regarded as a vector of random variables.
The CTMC capturing the model's evolution is based on the numerical state vector. At any given time, a joint probability distribution can be associated with this CTMC\@. 

Let $C_{u}$ be a local state of a sequential process in $\mathcal{C}_{\mathbb{M}}$.
The \emph{apparent rate} of an action type $\alpha$ in $C_{u}$, denoted by $r_{\alpha}(C_{u})$, is the total rate $C_{u}$ offers for $\alpha$ activities \cite{Hillston:1996}. Note that $C_{u}$ might have the form: 
\mbox{$C_{u}\!=\!(\alpha,r').C_{u}'\!+\!(\alpha,r'').C_{u}'' \!+\! \ldots$}
with multiple states reachable from $C_{u}$ via an $\alpha$ activity. 
\begin{eqnarray}
\nonumber r_{\alpha}(C_{u}) &\!\!\!\!\! =\!  	\left \{		\begin{array}{ll}
																					\!\!\!r, & \ \text{if} \ \ \  C_{u}\!=\! (\beta,r).P\ ,\ \beta=\alpha \\
																					\!\!\!0, & \ \text{if} \ \ \  C_{u}\!=\! (\beta,r).P\ , \ \beta\not=\alpha \\
																					\!\!\!r_{\alpha}(P) + r_{\alpha}(Q), & \ \text{if} \ \ \  C_{u}= P+Q\  \\
																			\end{array} \right. 
\end{eqnarray}
We define $r_{\alpha}(C_{u},C_{u}')$ to be the apparent rate of an action type $\alpha$ in  $C_{u}$ leading specifically to derivative $C_{u}'$. 
\begin{eqnarray}
\nonumber r_{\alpha}(C_{u},C_{u}') & \!\!\!\!\!=\!\!  	
\left \{		\begin{array}{ll}
																					&\!\!\!\!\!\!\!\! r,  \ \ \ \text{if} \ \  C_{u}\!=\! (\beta,r).P\! \wedge \!  \beta\!=\!\alpha\! \wedge\!  P\! =\! C_{u}'  \\
																					 &\!\!\!\!\!\!\!\!0,  \ \ \ \text{if} \ \  C_{u}\!=\! (\beta,r).P\! \wedge \!  ( \beta\not\!=\!\alpha\!\vee\! P \not\!=\! C_{u}')   \\
																					 &\!\!\!\!\!\!\!\!r_{\alpha}(P)\!+\!r_{\alpha}(Q), 	\ \ \ \ \  \ \ \text{if}  \ \  C_{u}= P+Q\   
																			\end{array} \right. 
\end{eqnarray}
Note that $C_u$ might be passive with respect to $\alpha$. 
Moreover, $C_u$ might enable more than one passive $\alpha$ activity. In such a case, each is given a weight ($\omega \in \mathbb{N}$), clarifying the relative probability assigned to these activities.  These are handled appropriately in the apparent rate calculations, but we omit details here due to space limitations.

Let us assume that in $H_i\in\group{M}$ some instances are in state $C_{u}\!\in\! ds^{*}(H_i)$, enabling $\alpha$ activities, where $\alpha\not\in\interface{M}{H_i}$. 
Undertaking the $\alpha$ activity by any of these instances causes $\xi(H_i,C_{u})$ to decrease by one and the number of instances in one of the one-step $\alpha$ derivatives to increase by one.
\vspace{-9pt}
\begin{multline*}
\!\!\!\!\langle \ \boldsymbol\xi(H_1), \ldots
 ,\boldsymbol\xi(H_i),\ldots, \boldsymbol\xi(H_{N^{h}_{\mathbb{M}}}) \rangle  \stackrel{(\alpha,R_{\alpha}(\boldsymbol\xi(H_i)))}{\xrightarrow{\hspace*{1.05cm}}} \\[-3pt]
 \langle\  \boldsymbol\xi(H_1),\ldots ,\boldsymbol\xi^{\prime}(H_i),\ldots,\boldsymbol\xi(H_{N^{h}_{\mathbb{M}}}) \ \rangle
\end{multline*}
Assume that the state change $C_{u} \stackrel{(\alpha,r_{\alpha})}{\longrightarrow} C_{u}^{\prime}$ occurs. Then:
\begin{eqnarray*}
\boldsymbol\xi(H_i)&\!\!\!\!\!\! = &\!\!\!\!\!\!\langle ...\,, \xi(H_i,C_{u}), ..\, ,\xi(H_i,C_{u}^{\prime}),..\, , \xi(H_i,C_{N_{H_i}}\!)\rangle  \\
\boldsymbol\xi^{\prime}\!(H_i) & \!\!\!\!\!\!= & \!\!\!\!\!\!\langle ...\,, \xi(H_i,C_{u})\!-\!1, ..\,, \xi(H_i,C_{u}')+1,..\, , \xi(H_i,C_{N_{H_i}}\!)\rangle 
\end{eqnarray*}
For $H\!\!\in\!\!\mathcal{G}(\mathbb{M}),C_{u},C_{u}' \in ds^{*}(H) $, $r^{*}_{\alpha}(H,C_{u},C_{u}')$ denotes the apparent rate of action type $\alpha$ in the context of $H$ restricted to activities which move from state $C_{u}$ to $C_{u}'$. 
\[ r^{*}_{\alpha}(H,C_{x,y},C_{x,y}^{\prime}) = \xi(H,C_{x,y}) \times r_{\alpha}(C_{x,y},C_{x,y}^{\prime})\]
According to the PEPA semantics, the rate of the transition above is: \mbox{$ R_{\alpha}(\boldsymbol\xi(H_i)) = \xi(H,C_{x,y}) \times r^{*}_{\alpha}(H_i,C_{u},C_{u}^{\prime}) $}.
We also define $r^{*}_\alpha(H)$, the apparent rate of $\alpha$ in $H$ offered by all instances in $H$; $r^{*}_{\alpha}(H)\!\!=\!\! \sum_{ C \in ds(sc(H)) }\! \xi(H,C)\!\times\! r_{\alpha}(C)$.
When instances in two distinct groups cooperate (say, $C_u$ instances in $H_i$ cooperate with $C_v$ instances in $H_j$ on a shared $\alpha$ activity becoming $C_u'$ and $C_v'$ respectively) the update to the numerical state vector will make changes analogous to those outlined above, i.e.\ the subvectors corresponding to the two groups.  The rate of the shared activity will be:
\begin{multline*}
R_{\alpha}(\boldsymbol\xi(H_i),\boldsymbol\xi(H_j)) = \\
\frac
{r^{*}_{\alpha}(H_i,C_{u},C_{u}')}
{r^{*}_{\alpha}(H_i)}
\frac
{{r^{*}_{\alpha}(H_j,C_{v},C_{v}')}}
{r^{*}_{\alpha}(H_j)} 
\min({r^{*}_{\alpha}(H_i)},{r^{*}_{\alpha}(H_j)})
\end{multline*}

%% file: AggregationCondition/new-aggregationCondition.tex
\section{Approximate Aggregation}
\label{condition}
\subsection{Syntactic Condition}
We consider a sub-class of large-scale PEPA models in which the populations of different sequential components have significantly different sizes. 
Consider a model $\mo{M}$ in this sub-class
and assume that according to their sizes,  $\mo{M}$'s groups are partitioned into two sets, large groups and small groups.
Let $\smgroup{M}$
denote the set of small groups in $\mo{M}$
and $\lagroup{M}$
denote its set of large groups:
$\group{M} = \smgroup{M}\  \cup \lagroup{M} $.
This set partition will also partition $\mathbb{M}$'s state vector --- the state variables  related to the groups in $\smgroup{M}$ and those related to the groups in $\lagroup{M}$. 
Without loss of generality, the model's state vector can be written as $\langle \boldsymbol\xi_{s},\boldsymbol\xi_{l}\rangle$ where $\boldsymbol\xi_{s}$ corresponds to the model's small groups, and  $\boldsymbol\xi_{l}$ to the large groups. 

As $\mathbb{M}$ is assumed to be a large-scale model, its complete CTMC is very large and its construction and analysis are computationally expensive. 
We show that if $\mo{M}$ satisfies the following condition, then we can perform an \emph{approximate} aggregation which results in an aggregated CTMC for $\mo{M}$.  We illustrate how this step enables us to efficiently estimate a marginal probability distribution over the model's small groups with a high accuracy. The cost of finding such a distribution using the aggregated CTMC is orders of magnitude smaller than through the analysis of the original CTMC\@. 
\begin{condition}
\label{aggregationCondition}
{\bf Syntactic Aggregation Condition} For any shared activity, synchronised between one or more large groups and one or more small group, the rate of the shared activity, should be completely decided by the small groups:
\begin{multline*}
\!\!\!\! \forall\ H^{l} \!\!\in\! \lagroup{M}\ \  \forall\  \alpha\!\in\! \einterface{M}{H} \\[-3pt]
 [\ \   (\ \coop{M}{H^{l}}{\alpha}\ \cap\ \smgroup{M} \ ) \not= \emptyset  \implies \quad \quad \\[-3pt]
 \ \ \ \ \ \ \ \forall\ C_{x,y}\  \in\ ds^{*}(H^{l})\ ,\ \exists\  \omega \in \mathbb{N}\ , \ \ r_{\alpha}(C_{x,y}) = \omega \top \ ]
\end{multline*}
\end{condition}
The condition expresses that in any synchronization on a shared activity of type $\alpha$, if both small and large groups are involved, then all instances in the involved large groups need to undertake $\alpha$ passively. 
Consider a variant of the client-server model with the client components modified as:
\begin{eqnarray*}
\ \ C_{\textit{think}}  \rmdef (\textit{think}, r_t).C_{\textit{req}}\ \ \ \ \ \ \  \ \ \ C_{\textit{req}}  \rmdef  (\textit{req}, \top).C_{\textit{think}}
\end{eqnarray*}
We assume $\textit{Clients}$ constitutes a large group, and $\textit{Servers}$, a small one.  They synchronise on the $\textit{req}$ activity which the clients undertake passively.  Thus the model now satisfies Condition~1.
We will see that this implies that in any state of the system, the global apparent rate of $req$ depends only on the configuration of the servers and is independent of the clients' configuration.

\subsection{The CTMC Structure}
\label{sec:ctmcstructure}

The complete CTMC of a large-scale model with small groups which satisfies Condition \ref{aggregationCondition} exhibits important structural properties which can be exploited in order to build an aggregated CTMC\@. 
In this section, we introduce these properties and illustrate them in the CTMC of the modified client-server model initialised with two servers and two clients. 

\begin{figure*}[t]
\begin{subfigure}[t]{0.85\textwidth}
                \includegraphics[width=1\textwidth,height=0.40\textheight]{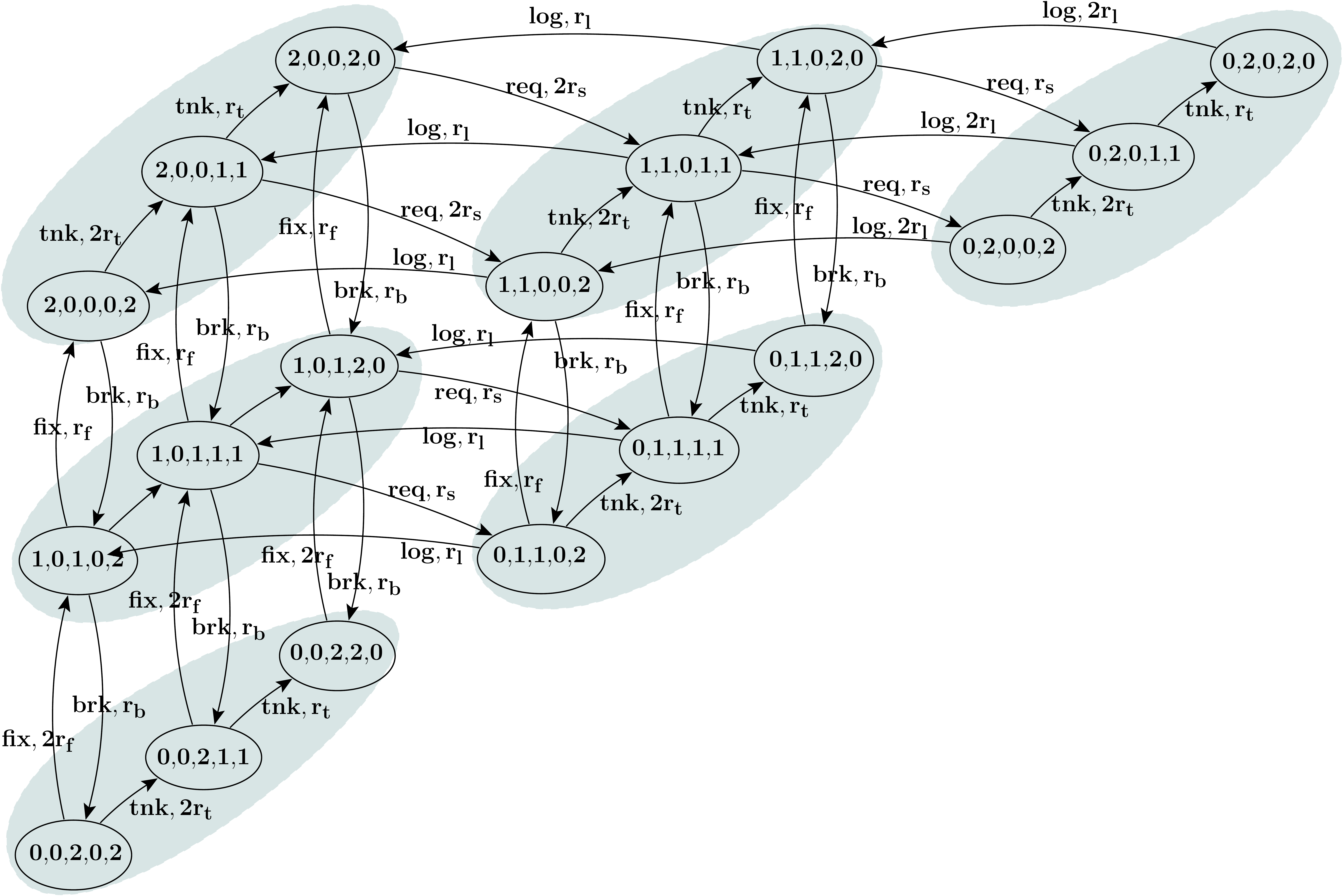}
                \caption{Complete CTMC.}
                \label{fig:fullstatesapce}
\end{subfigure}
\!\!\!\!\!\!\!\!\!\!\!\!\!\!\!\!\!\!\!\!\!\!\!\!\!\!\!\!\!\!\!\!\!\!\!\!\!\!\!\!\!\!\!\!\!\!\!\!\!\!\!\!\!\!\!\!\!\!\!\!\!\!\!\!\!\!\!\!\!\!\!\!\!\!\!\!\!\!\!\!\!\!\!\!\!\!\!\!\!\!\!\!\!\!\!
\begin{subfigure}[t]{0.10\textwidth}
                \includegraphics[width=3.1\textwidth,height=0.19\textheight]{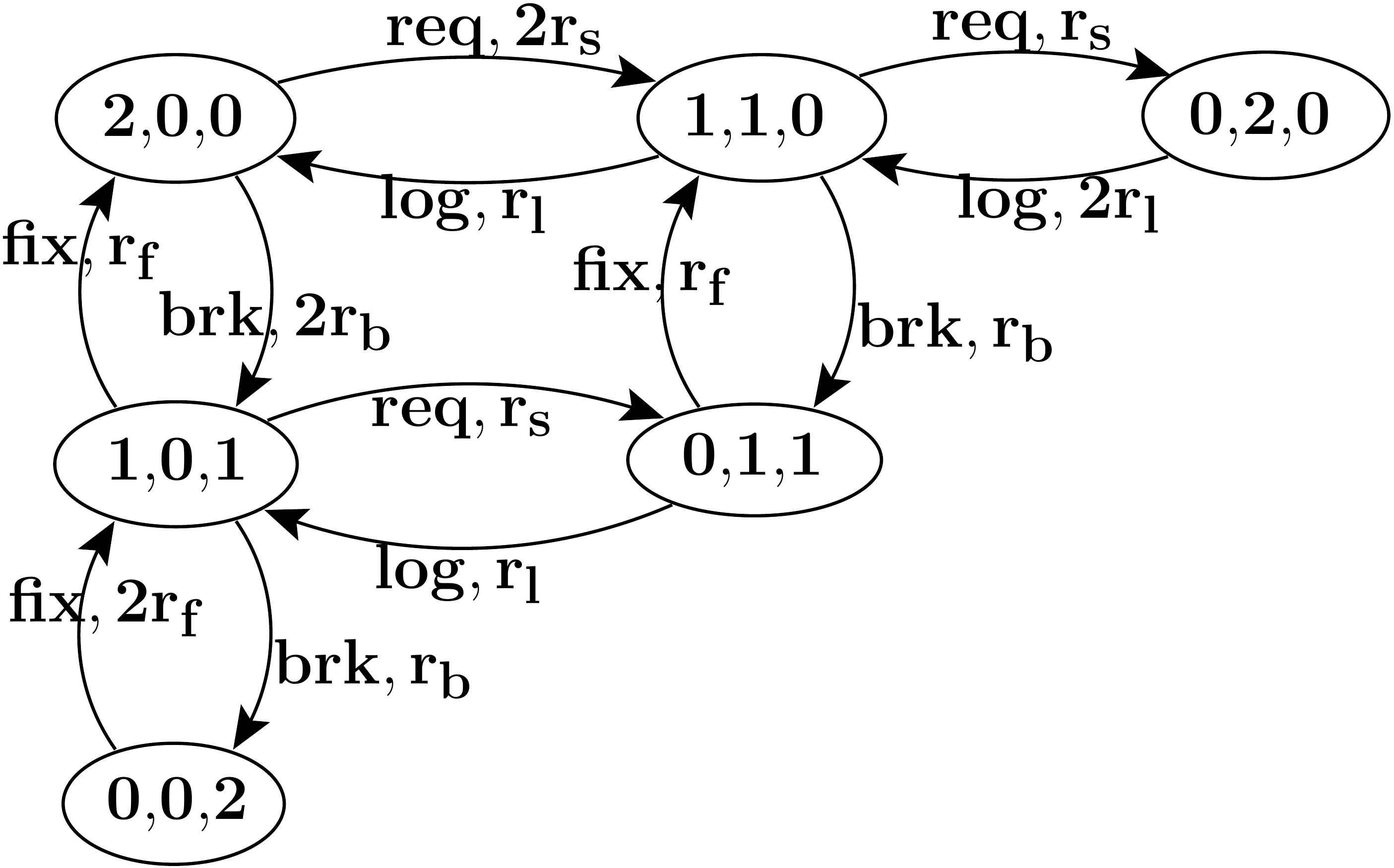}
						\caption{Aggregated CTMC.\label{fig:aggStateSpace}}
\end{subfigure}
\caption{Complete and aggregated CTMC for the client-server system with two servers and two clients (shorthand $tnk=think$ is used). Each state of the complete CTMC is a vector $\langle S_i,S_l,S_b,C_r,C_t\rangle$, where $S_i$ counts the number of idle servers, $S_l$ the number of logging servers, $S_b$ the number of broken servers, $C_r$ the number of requesting clients and $C_t$ the number of thinking clients. Each state of the aggregated CTMC follows this state descriptor: $\langle S_i,S_l,S_b\rangle$.}
\end{figure*}

Given that the model's groups are partitioned by size, we can categorise each transition depending on whether it changes the state of only one or more small groups, the state of only one or more large groups  or simultaneously, the state of both small and large groups. 
For $H\in\group{M}$, let us define \mbox{$\actions{H}$} to be the set of action types enabled by $H$ and \mbox{$\actionsI{H}=\actions{H}-\interface{M}{H}$} represents the set of action types 
of individual activities undertaken by instances in $H$. 
First, assume that $H\!\!\in\!\!\lagroup{M}$ and define \mbox{$\actionsLargeGroup{H}$ to} be the set of action types instances in $H$ undertaken individually or in cooperation with other large groups.
\begin{multline*}
\actionsLargeGroup{H}\!=\!\! \{\ \alpha \mid\alpha\! \in \actionsI{H}\ \vee\  
 (\coop{M}{H}{\alpha}\  \cap\  \smgroup{M})\!=\! \emptyset \ \}
\end{multline*}
Based on $\actionsLargeGroup{H}$, we define $\actionsLargeGroup{\mo{M}} = \bigcup_{H \in \lagroup{M}}^{} \actionsLarge{H}$, the set of action types related to the dynamics of the large groups only. 
As the second case, assume that $H\!\in\!\smgroup{M}$ and define $\actionsSmallGroup{H}$ analogously for small groups: 
\begin{multline*}
\actionsSmallGroup{H}\!=\!\! \{\ \alpha \mid\alpha\! \in \actionsI{H}\ \vee\  
 (\coop{M}{H}{\alpha}\  \cap\  \lagroup{M})\!=\! \emptyset \ \}
\end{multline*}
As above, we define $\actionsSmallModel{M}= \bigcup_{H \in \smgroup{M}}^{} \actionsSmallGroup{H}$ which represents the action types in which the large groups do not participate. 
Finally, for $H\in\smgroup{M}$, we define $\actionsSmallLargeGroup{H}$ to represent the set of action types shared between instances of $H$ and instances of one or more large groups. 
\[\actionsSmallLargeModel{M} = \{\ \alpha\mid\alpha\in\interface{M}{H} \wedge\ (\coop{M}{H}{\alpha}\ \cap\ \lagroup{M})\not\!=\!\emptyset\}\]
Similarly, for model $\mo{M}$, $\actionsSmallLargeModel{M} = \bigcup_{H\in\smgroup{M}}^{}\ \actionsSmallLargeGroup{H}$ is defined to be the set of action types in which both small and large groups participate.  
This categorisation of the action types will be used to characterise the structural properties exhibited by the CTMC of the models satisfying Condition~\ref{aggregationCondition}.
For the modified client-server system, $\actionsLargeModelNormal{CS}\!\!=\!\!\{\textit{think}\}$, $\actionsSmallLargeModelNormal{CS}\!\! =\!\! \{\textit{req}\}$ and $\actionsSmallModelNormal{CS}\!\!=\!\! \{\textit{log,break,fix}\}$.
%

\subsubsection{Identifiable Sub-Chains}
Consider a model $\mo{M}$ satisfying Condition~\ref{aggregationCondition}. 
In its underlying CTMC, let us focus on transitions of type $\actionsLargeModel{M}$.
These transitions only change the state of one or more large groups, leaving the configuration of the small groups unchanged.  Based on this, the CTMC can be divided into sub-chains where the states within each sub-chain are connected by $\actionsLargeModel{M}$ transitions and for which the configuration of instances in the small groups remains the same. 
From a state $S$, a sub-chain $Y_i$, 
can be derived using these rules: 
\begin{eqnarray*}
&& S \in Y_i \ \wedge\ S \stackrel{(\alpha,\cdot)}{\longrightarrow} S^{\prime}\  \wedge\  \alpha \in \actionsLargeModel{M} \implies S^{\prime} \in Y_i \\
&& S \in Y_i \ \wedge\ S^{\prime\prime} \stackrel{(\alpha,\cdot)}{\longrightarrow} S\  \wedge\  \alpha \in \actionsLargeModel{M} \implies S^{\prime\prime} \in Y_i
\end{eqnarray*}
The sub-chain $Y_i$ associated with a state $S_i$ consists of $S_i$, the states from which $S_i$ can be reached by $\actionsLargeModel{M}$ transitions, and the states reachable from $S$ by such transitions. These transitions capture the dynamics of the large groups in $Y_i$. 

Using the rules above, given 
$\smgroup{M}$ and $\lagroup{M}$, 
for $\mo{M}$'s underlying CTMC, the partition $Y_{\mo{M}}\!\!=\{Y_1,Y_2,\ldots\}$ can be formed where each $Y_i$ is a sub-chain which
can be uniquely identified by the configuration it captures for $\mo{M}$'s small groups (the vector $\boldsymbol{\xi}_{s}$). Two states $S_1$ and $S_2$ in the original CTMC belong to two different sub-chains if they do not exhibit the same configuration for instances of groups in $\smgroup{M}$.

In Fig. \ref{fig:fullstatesapce}, the CTMC of the modified model, the sub-chains are visually identified.
In each sub-chain, the clients change their state via $\textit{think}$ activities without affecting the servers. For a model with a larger client population,  the same partition, but with longer sub-chains, would be observable.

\subsubsection{Unlikely Boundary States, Blocked Activities}
Before showing the next structural property, we need to define the notion of boundary states. Within the CTMC of $\mo{M}$, consider an arbitrary sub-chain $Y_i$ and $\alpha \in \actions{M}$. 
First, let us assume that $\alpha \in \actionsLargeGroup{M}$, i.e.\ small groups do not participate in $\alpha$ activities. 
In this case, in any state $S\in Y_i$, whether $S$ enables $\alpha$ activities or not, depends only on the configuration of the large groups in $S$. 
As the second case, assume that $\alpha\in \actionsSmallModel{M}$.
Here, the large groups do not participate in the $\alpha$ activities and therefore, in any state $S\in Y_i$, whether $\alpha$ is enabled depends only on the configuration $S$ captures for the small groups. 
Moreover, since the configuration of the small groups is identical across all the states in $Y_i$, the status of $\alpha$ is the same across $Y_i$ as a whole. 

The third case is when $\alpha \in \actionsSmallLargeModel{M}$. Here, $\alpha$ activities are shared between one or more small and one or more large groups and the status of $\alpha$ in any state depends on the configurations of groups of both types. Here we can identify a subset of states in $Y_i$, where shared $\alpha$ activities are not enabled because there are no instances in one or more relevant large groups to participate.  Extending this to all action types in $\actionsSmallLargeModel{M}$, in each sub-chain $Y_i$, we can find a subset of states where one or more action types $\alpha$ remain disabled specifically due to the lack of cooperation from the instances in the large groups. For $Y_i$, let us refer to this subset as the \emph{boundary states} of $Y_i$, denoted $bo(Y_i)$. 
For instance, in the client-server example, the boundary states are those where there are no clients ready to make a request. 
The formal characterisation of the boundary states follows: 
\begin{mydef}
\label{blockedstate}
In $\mo{M}$'s state space, a state $S$ with the vector $\boldsymbol\xi\!=\! \langle\boldsymbol\xi(H_1),\ldots,\boldsymbol\xi(H_{N_{\mo{M}}^{H}})\rangle$, is a boundary state if and only if:
\begin{multline*}
\!\!\! \ \exists\  H_s \in \smgroup{M} \ \  \exists\  \alpha \in (\actions{H_s}\  \cap \actionsSmallLargeModel{M})\ \\[-2pt]  \ \ \exists\  H_l \in (\lagroup{M} \cap \coop{M}{H_s}{\alpha})\ : 
\ \ \ r_{\alpha}^{*}(H_{l})=0 
\end{multline*} 
\end{mydef}

We define $bl(Y_i,\alpha)$, $Y_i\!\in Y_{\mo{M}},\alpha\! \in \actionsSmallLargeModel{M}$, to represent the set of states in $Y_i$ where $\alpha$ is blocked. 
\begin{eqnarray*}
bl(Y_i,\alpha)&\!\!\!\! = &\!\!\!\left\{ S \mid S \in Y_i \ \wedge\ \alpha \ \in bl'(S)  \right\} \\[-2pt] 
\!bl'(S)&\!\!\!\!=& \!\!\!\!\!\left\{\alpha \mid \exists H_s \in\smgroup{M} : \left[ \  \alpha \in (\actions{H_s}\,\cap\ \actionsSmallLargeModel{M}) \ \wedge \right. \right.   \\[-3pt]
&& \!\!\!\!\!\!\!\!\!\!\!\!\left.\left. \exists\  H_l \in ( \lagroup{M}\,\cap \coop{M}{H_s}{\alpha} ) : r_{\alpha}^{*}(H_l)=0 \vphantom{\actionsSmallLargeModel{M}} \ \right]  \ \ \right\} 
\end{eqnarray*}
In large-scale resource-bound systems, resources are unlikely to remain idle waiting for the resource users to ask for the service. Therefore, in such systems, the probability of being in the boundary states can be assumed to be negligible.

%
\subsubsection{Rate Regularities}
\begin{figure*}[t]
\centering
\begin{subfigure}[b]{0.45\textwidth}
							\centering 
                \includegraphics[width=0.7\textwidth,height=0.13\textheight]{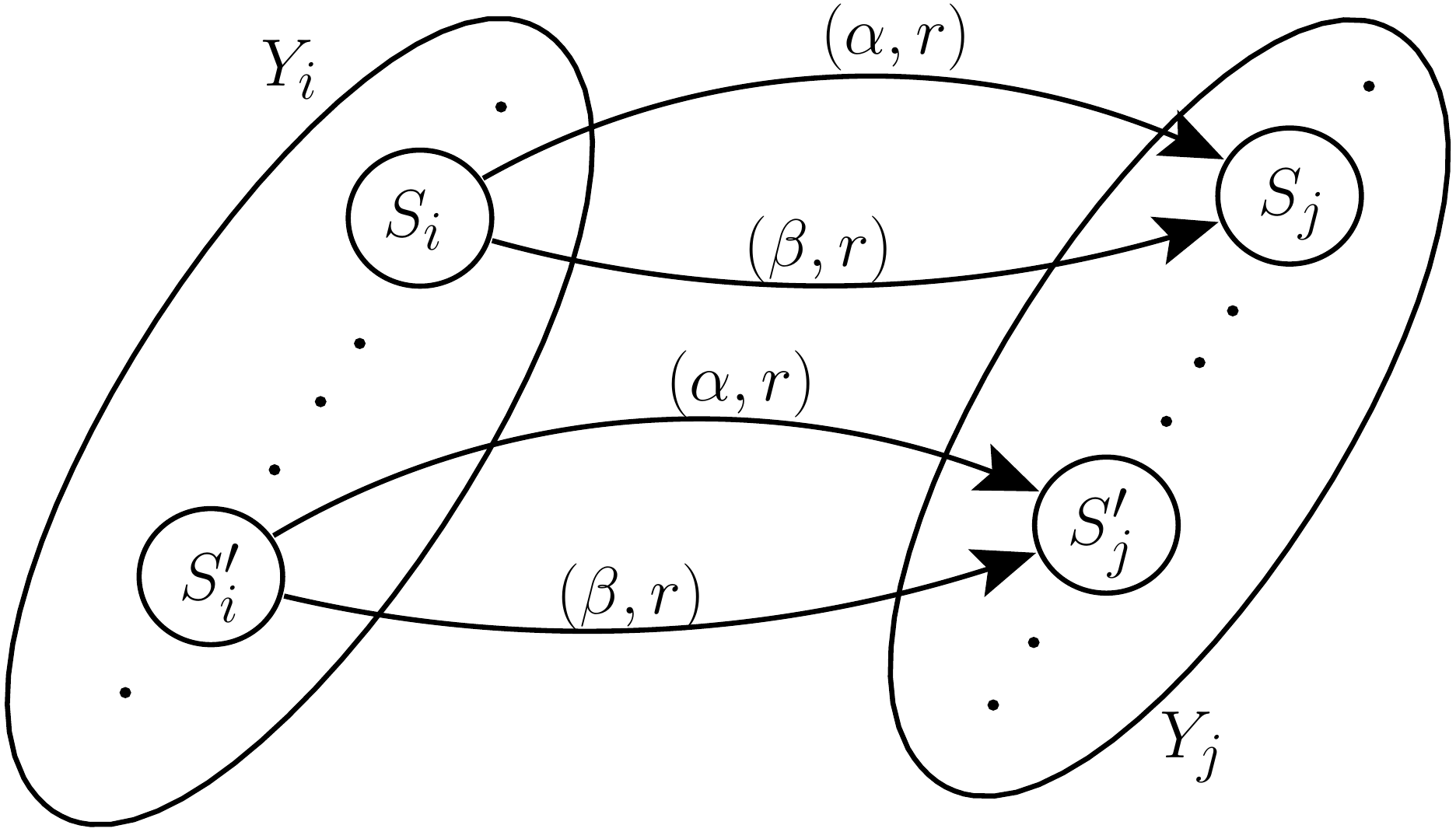}
                \caption{The non-boundary states in a sub-chain presenting similar behaviours.}
						\label{fig:similarBehaviours}
\end{subfigure}
\begin{subfigure}[b]{0.45\textwidth}
						\centering 
						\includegraphics[width=0.7\textwidth,height=0.13\textheight]{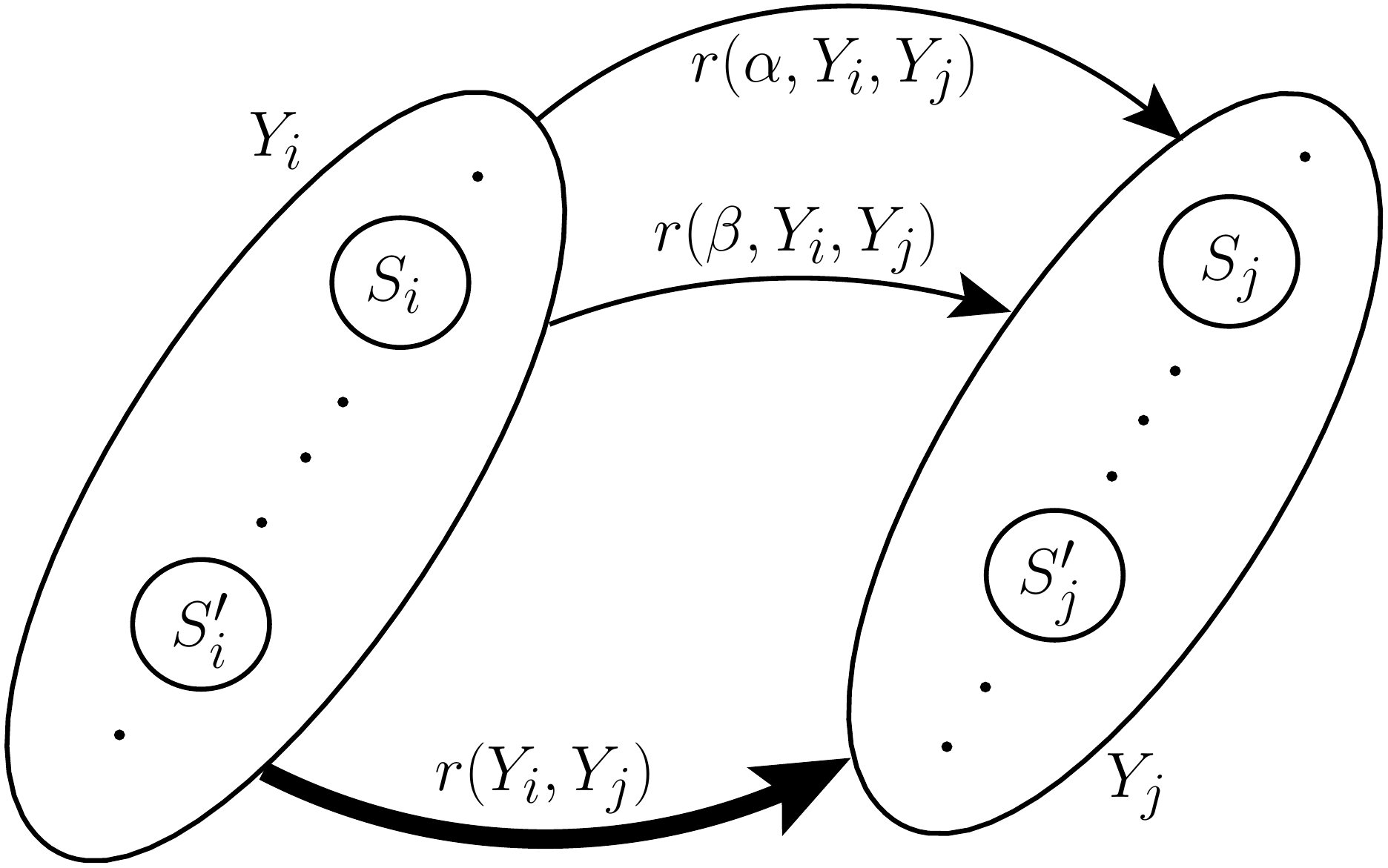}
						\caption{The rate for any one cross-sub-chain transition.}
						\label{}
\end{subfigure}
\caption{Rate regularities for cross-sub-chain transitions.}
\end{figure*}

Consider $\mo{M}$, a model satisfying Condition~\ref{aggregationCondition}. In the underlying CTMC, consider a sub-chain $Y_i$ and state $S_i\! \in\! Y_i$. 
Assume that there exists a cross-sub-chain transition $S_i\!\! \stackrel{(\alpha,R)}{\longrightarrow}\!\! S_j$ where $S_j\!\! \in\! Y_j,Y_j\!\! \not=\! Y_i$ (see Fig.~\ref{fig:similarBehaviours}). 
Then for any $\alpha$ enabling state $S_i^{\prime}\!\! \in\! Y_{i}$, we can observe a similar transition $\transition{S_i^{\prime}}{S_j^{\prime}}{\alpha,R^{\prime}}$ where $S_j^{\prime}\! \in\! Y_j$ and $R\!\!=\!\!R^{\prime}$.  This observation leads to the following Lemma.
\begin{lemma} Rate regularity for non-boundary states.
\label{observation}
\begin{multline*}
\forall\  \alpha \in (\actionsSmallLargeModel{M}\ \cup\ \actionsSmallModel{M})\ ,\ \  \forall\   S_i\  \in\  Y_i\ ,\ \  \forall\  S_j\  \in\  Y_j  \\[-2pt]
 \left(\  \transition{S_i}{S_j}{\alpha,R} \implies 
\forall\ S_i^{\prime}\ \in\ (Y_i - bl(Y_i,\alpha)) \right. \\[-2pt]
 \ \ \ \left. \left[\  r_{\alpha}^{*}(S_i^{\prime}>0) \implies \exists\ S_j^{\prime}\  \in\  Y_j\ \ \ \transition{S_i^{\prime}}{S_j^{\prime}}{\alpha,R} \ \right]\  \right)
\end{multline*}
\end{lemma}
In any sub-chain, all non-boundary states enable the same set of activities.
Hence, for $\alpha\!\! \in \actionsSmallLargeModel{M} \cup\!\!\actionsSmallModel{M}$, we can define $r(\alpha,Y_i,Y_j)$, the rate of any $\alpha$ transition from any $S_i\in\! Y_i$ to another state $S_j \in\! Y_j$.
\begin{eqnarray*}
r(\alpha,Y_i,Y_j) &\!\!\! =\!\!\! & \left\{
\begin{array}{ll}
r & \quad \text{if}\ \ \ \exists\ S_i \in Y_i\  \exists\  S_j \in Y_j : \  \transition{S_i}{S_j}{\alpha,r} \\
0 & \quad \text{otherwise} 
\end{array} \right . 
\end{eqnarray*}
Between any two sub-chains $Y_i$ and $Y_j$, transitions of more than one type might connect their states (see Fig. \ref{fig:similarBehaviours}). We define $r(Y_i,Y_j)\!=\! \sum_{\alpha\in(\actionsSmallModel{M}\ \cup\  \actionsSmallLargeModel{M})}^{}\ r(\alpha,Y_i,Y_j)$
 to represent the total rate at which non-boundary states in $Y_i$ transition into a corresponding state in $Y_j$. 

In Fig.~\ref{fig:fullstatesapce}, the activities of type $\actionsSmallModelNormal{CS}\ \cup\!\actionsSmallLargeModelNormal{CS} = \{\textit{req,log,break,fix}\}$ cause the system to leave the sub-chain it resides in and move into a new one. The rate of these activities depends only on the configuration of the servers. For the states within one sub-chain, the state of the servers does not change. Thus, the rate of any activity of type $\actionsSmallModelNormal{CS}\cup\!\actionsSmallLargeModelNormal{CS}$, if enabled, is the same. 

For a large-scale model $\mo{M}$ that satisfies Condition \ref{aggregationCondition}, we can construct its CTMC and use $\actionsSmallModel{M}$, $\actionsLargeModel{M}$, and $\actionsSmallLargeModel{M}$ to detect the sub-chains formed and observe the rate regularities for transitions between sub-chains. 
These regularities and the assumption that the probability of experiencing the boundary states is negligible, enable us to build an aggregated CTMC for the model. 
In this CTMC, each sub-chain is represented by a single aggregate state. This aggregate state abstracts away the dynamics of the large groups in that sub-chain and only captures the un-changed configuration the sub-chain exhibits for the small groups. In other words, the aggregated CTMC captures only the evolution of the small groups and information about the behaviour of the instances in the large groups is lost. Nevertheless,
the aggregate CTMC provides the means to study the evolution of the instances of the groups in $\smgroup{M}$ in a fined-grained manner without building the very large original CTMC. 

Fig. \ref{fig:aggStateSpace} shows the aggregated CTMC of the modified client-server model. It shows how the state variables related to the servers behave.

%% file: Algorithm/new-algorithm.tex
\section{Aggregation Algorithm}
\label{sec:aggregationAlgorithm}

The aggregated CTMC of a conforming model $\mo{M}$ is more abstract than its original CTMC (capturing only the state and transitions of the instances in $\smgroup{M}$) and its structure is independent of the size of the large group populations.
This independence justifies the development of an algorithm which builds the  aggregated CTMC \emph{directly} from the model specification.
In a first reduction step, our algorithm transforms the system equation of the original model $\mo{M}$ into $\mo{M}_{R}$, a \emph{reduced} form which captures $\mo{M}$'s structure only with respect to the groups in $\smgroup{M}$. 
In the next step, using $\mo{M}_{R}$ and a semantics developed for PEPA population models, the aggregated CTMC is generated. These steps are described in this section. 
Note that the reduced system equation faithfully captures the synchronization restrictions imposed on any of the small groups. Although the large groups are removed from the model, the reduction rules guarantee the behaviour exhibited considering $\mo{M}_{R}$ matches the one seen for $\smgroup{M}$ in the \mbox{original CTMC}.

\subsection{Reduction}
These rules will be applied to the system equation of an input model $\mo{M}$ in order to produce the reduced model $\mo{M}_R$. 
\begin{eqnarray*}
red({G}) \!&\!\!\! = \!\!\! &\!\!\! 
\left\{\begin{array}{ll}
											\!\!\!\!red({G}_1) \mysync{L} red({G}_2), & \text{if}\ {G}\! =\! {G}_1  \mysync{L}  {G}_2  \\[2pt]
											\!\!\!\!H\{\cdot\}, & \text{if}\ {G}=H\{\cdot\}\  , \  H \in \smgroup{M}	\\[2pt]
											\!\!\!\!Nil, & \text{if}\ {G}=H\{\cdot\} \ , \  H \in \lagroup{M}	\\[2pt] 
											\!\!\!\!red(X), & \text{if}\ \ {G} \rmdef X 
										\end{array} \right. 
\end{eqnarray*}
The process $Nil$ represents a sequential process which does not undertake any activity. Applying the rules results in a system equation where all large groups are replaced by this process. The following rules remove $Nil$ processes to find the minimal reduced system equation:
\begin{equation*}
Nil \mysync{\cdot} Nil = Nil \quad \quad
Nil \mysync{\cdot} P = P \quad \quad P \mysync{\cdot} Nil = P
\end{equation*}
\subsection{Semantics}
\label{semanticsSection}

Having built the reduced form of $\mo{M}$'s system equation, we apply the \emph{count-oriented structured operational semantics} presented below to derive the model\rq{}s underlying labelled transition system (LTS) \emph{directly} in numerical vector form. 
\begin{framed}
\textbf{Promotion}. Promotion of a sequential component's transition to the group level:
\begin{eqnarray*}
\!\!\!\!\!
\setpremisesend{0pt}
\inference []{C_{u} \stackrel{(\alpha,r_{\alpha})}{\longrightarrow} C_{u}'}
{\boldsymbol\xi(H)\!\stackrel{(\alpha,\xi(H,C_{u})r_{\alpha}(C_{u},C_{u}'))}{\verylongsquiggly}\Theta(\boldsymbol\xi(H),C_{u},C_{u}')} \ \xi(H,C_{u})\!>\!0
\end{eqnarray*}
\\[6pt]
\textbf{Cooperation}. Cooperation between groups:
\begin{eqnarray*}
&&\!\!\!\!\!\!\!\!\!\!\!\!\!\!\!\!
\inference[]{\boldsymbol\xi(H_i) \stackrel{(\alpha,r(\boldsymbol\xi(H_i)))}{\longsquiggly} \boldsymbol\xi^{\prime}(H_i) }
{\boldsymbol\xi(H_i) \mysync{L} \boldsymbol\xi(H_j)\stackrel{\alpha,r(\boldsymbol\xi(H_i)))}{\longsquiggly} \boldsymbol\xi^{\prime}(H_i)\mysync{L}\boldsymbol\xi(H_j)}\ \alpha\! \not \in\! L 
\\[4pt]
&&\!\!\!\!\!\!\!\!\!\!\!\!\!\!\!\!
 \inference[]{\!\!\!\boldsymbol\xi(H_i)\!\! \stackrel{(\alpha,r_1(\boldsymbol\xi(H_i)))}
{\longsquiggly} \!\!\boldsymbol\xi^{\prime}(H_i) \ \wedge\  \boldsymbol\xi(H_j)\!\! \stackrel{(\alpha,r_2(\boldsymbol\xi(H_j)))}{\longsquiggly} \!\!\boldsymbol\xi^{\prime}(H_j)}{\boldsymbol\xi(H_i) \mysync{L} \boldsymbol\xi(H_j)\stackrel{(\alpha,R)}{\longsquiggly} \boldsymbol\xi^{\prime}(H_i)\mysync{L}\boldsymbol\xi^{\prime}(H_j)}  \alpha\! \in\! L  
\\[4pt]
&&\!\!\!\!\!\!\!\!\!
R = \frac{r_1(\boldsymbol\xi(H_i))}{r^{*}_{\alpha}(H_i)} \frac{r_2(\boldsymbol\xi(H_j))}{r^{*}_{\alpha}(H_j)} min(r^{*}_{\alpha}(H_i),r^{*}_{\alpha}(H_j)) 
\end{eqnarray*}

\end{framed}
\begin{framed}
\textbf{Constant}. Process constants: 
\begin{eqnarray*}
&&\inference []
{M \stackrel{(\alpha,r)}{\longsquiggly} M^{\prime}} 
{A \stackrel{(\alpha,r)}{\longsquiggly} M^{\prime}}\ A \rmdef M
\end{eqnarray*}
\end{framed}
Due to  the space limitation, the symmetric rule for the first cooperation rule (evolution of the right hand process on individual activities) is omitted. 
The rules for constructing the transition relation~$\shortsquiggly$ make use of the transition relation $\rightarrow$ built by the original PEPA semantics \cite{Hillston:1996} (see the premise of the first rule). However, note that $\rightarrow$ needs to be constructed only at the level of the sequential processes; for each sequential process in the model an automata is constructed which captures the states and transitions each instance of that sequential process experiences.

The semantics formally reflects the update to the numerical vector representation of the states, as described in Section~2, recording the impact of the completion of an action on each of the counts for the involved groups.  The appropriate update is determined by the function $\Theta$.
\begin{eqnarray*}
&& \!\!\!\!\!\!\!\!\!\!\!
\Theta(\boldsymbol{\xi}(H),C_{u},C_{u}') \!\!= \!\! \langle \Theta^{\prime}(\boldsymbol{\xi}(H),C_{u},C_{u}',\xi(H,C_{u}^{\prime\prime}))\!\mid\! C_u^{\prime\prime}\!\in\!ds^{*}\!(H)\rangle \\
&& \!\!\!\!\!\!\!\!\!\!\!
\Theta^{\prime}(\boldsymbol{\xi}(H),C_{u},C_{u}',\xi(H,C_u^{\prime\prime})\!\! =\!\! \left\{
\begin{array}{ll}
\xi(H,C^{\prime\prime})-1, &  \ \ \ \text{if}\ C^{\prime\prime}\!\! =\!C_{u}\\
\xi(H,C^{\prime\prime})+1, &  \ \ \ \text{if}\  C^{\prime\prime}\!\! =\! C_{u}'\\
\xi(H,C^{\prime\prime}),	 &  \ \ \ \text{otherwise}\ \ \ 
\end{array}\right.
\end{eqnarray*}
\subsection{Generating the Aggregate CTMC}
In the aggregation algorithm the count-oriented semantics is applied to the reduced form system equation of a model, creating its aggregated LTS\@. For a model $\mo{M}$, this LTS, which is formally characterised below, constitutes the model's aggregated CTMC\@.

\textbf{Aggregated Derivative Set}. The aggregated derivative set of a state $S$, denoted by $D^{*}(S)$, is the smallest set of states satisfying the following conditions. 
\begin{eqnarray*}
&& 1. \ \ S \in D^{*}(S). \\[-3pt]
&& 2. \ \ \text{If}\ \  S_1 \in D^{*}(S)\  \text{and} \ S_1 \stackrel{(\alpha,\cdot)}{\shortsquiggly} S_2, \  \text{then}\  S_2 \in D^{*}(S).
\end{eqnarray*}

\textbf{Aggregated Labelled Transition System} is defined as a tuple \mbox{$(D^{*}(S_0),\Omega,\!\!\shortsquiggly\!\!\!\!)$}. $S_0$ is the model's initial aggregate state. $\Omega = (\mathcal{A} \times \mathcal{F} ) $ is the transition system's alphabet where $\mathcal{A}$ is the set of action types defined in $\mo{M}$ and $\mathcal{F}$ is a function space defined over $\mathbb{Z}_{+}^{d_{\mo{M}_R}}$. $d_{\mo{M}_R}$ is the dimension of the reduced system equation, i.e.\ the number of state variables appearing in its underlying state vector. Each function in $\mathcal{F}$ corresponds to an action type: $F_{\alpha}$ defines the rate of observing an activity of type $\alpha$ in the vector space $\mathbb{Z}_{+}^{d_{\mo{M}_R}}$. 

Fig.~\ref{fig:aggStateSpace} shows the aggregated CTMC constructed for the modified Model~1 when initialised with two idle servers.

%% file: CalculatingMarginals/new-calculatingMarginals.tex
\section{\!\!\!\!\!Deriving Marginal distributions}
\label{equationAlgorithm}
\textbf{Derivation through the aggregated CTMC}. 
The numerical state vector of a PEPA model can be regarded as a vector of dependent random variables whose evolution is captured by the underlying CTMC\@.
For this CTMC, the Chapman-Kolmogorov (C-K) equations (a set of linear differential equations) can be derived for the evolution of the joint probability distribution of the state random variables over time.
For any model, the number of C-K equations is equal to the size of the state space of the CTMC\@.
Therefore, as explained earlier, calculating the full joint probability distributions for large-scale PEPA models is usually infeasible. 

Recall that the state vector $\boldsymbol\xi$ of a large-scale PEPA model $\mo{M}$ satisfying the aggregation conditions can be expressed as $\boldsymbol\xi\! =\! \langle \boldsymbol\xi_{s}, \boldsymbol\xi_{l}\rangle$ where $\boldsymbol\xi_{s}$ captures the state of instances within groups in $\smgroup{M}$ and $\boldsymbol\xi_l$, the state of the instances within groups in $\lagroup{M}$. 
The aggregation algorithm constructs an aggregated CTMC in which only the evolution of  $\boldsymbol\xi_{s}$ is captured. 
This CTMC can be used to derive \emph{marginal} probability distributions over the random variables $\xi(H,C_{x,y}),\forall H_i\! \in\! \smgroup{M}, \forall C_{x,y}\! \in\! ds\!^{*}(H_i)$. These are calculated by a set of ordinary differential equations (ODE). In the following, assume that $X_a$ denotes the stochastic process whose behaviour is captured by the aggregated CTMC\@. For each state of this CTMC, one equation is constructed: 
\begin{multline*}
\!\!\!\!\!\! \frac{d\; \mathbb{P}(X_a\! =\! S)}{d\ t}\!=\sum_{\alpha:S_j\!\!\stackrel{(\alpha,r_{\alpha}^{*}(\boldsymbol\xi))}{\longsquiggly}\! S} \!\!\!\!\!\!\! r^{*}_{\alpha}(\boldsymbol\xi) \; \mathbb{P}(X_a\!=\!S_j) 
 \\[-3pt]
 - \!\!\!\!\!\!\!\!\sum_{\alpha:S\!\! \stackrel{(\alpha,r_{\alpha}^{*}(S))}{\longsquiggly}\! S_j} \!\!\!\!\!\!\!\!\; r_{\alpha}^{*}(S) \; \mathbb{P}(X_a\!=\!S)\;
\end{multline*}
The algorithm in Program.~\ref{alg:generatingODEs} constructs this system of ODEs.
\mathligsoff
\begin{program}[t]
\begin{alltt}
for each aggregate state \(S\) in \(ds\sp{*}(S\sb{0})\) form one 
equation \(\frac{d \mathbb{P}(X\sb{a}=S)}{dt}\) as \(E\) 

\(in\sb{S} \leftarrow in(S) = \{S\sb{h} | S\sb{h}\stackrel{(\alpha,\cdot)}{\shortsquiggly}S \} \)
\(out\sb{S} \leftarrow out(S) = \{S\sb{j} | S\stackrel{(\alpha,\cdot)}{\shortsquiggly}S\sb{j}\}\)      
    for each state \(S\sb{h}\!\in\!in\sb{S}:S\sb{h}\!\stackrel{(\alpha,r\sb{\alpha}\sp{*}(\boldsymbol\xi))}{\longsquiggly}\!S \)    
  						  												 //probability flux into \(S\)
\ \ \ \ \ \ \ add  \( + r\sb{\alpha}\sp{*}(\boldsymbol\xi) \times \mathbb{P}(X\sb{a}=S\sb{h}) \)\ to\ \(E\)    

    for each state \(S\sb{j}\!\in\!out\sb{S}:S\!\stackrel{(\alpha,r\sp{*}\sb{\alpha}(\boldsymbol\xi))}{\longsquiggly}\!S\sb{j} \) 
 						  												 //probability flux out of \(S\)
        add  \(-\ r\sp{*}\sb{\alpha}(\boldsymbol{\xi}) \times \mathbb{P}(X\sb{a}=S)\)   to \(E\)    
\end{alltt}
\caption{Pseudo-code that generates the ODEs for a model's marginal probability distributions.\label{alg:generatingODEs}}
\end{program} 
\mathligson

\textbf{Derivation of the marginal distribution from the \mbox{C-K} equations}.
For a PEPA model, its system of C-K equations can be constructed by forming an equation for each state of the model, recording the flux in and the flux out of that state. This allows analysis at the level of individual states.
In this section, we show a high level overview of how it is possible to construct the equations for evaluating the probability of being in each sub-chain using the model's original C-K equations. 

%
Let $X$ be the random process captured by $\mo{M}$\rq{}s original CTMC and $\mathbb{P}_{\langle X=S_i\rangle}(t)$ denote the probability that at time $t$, $X$ is in state $S_i$. In the model's C-K equations, one equation is constructed for each state $X$ experiences: 
\begin{multline}
\label{simpleCKEquation}
\!\!\!\!\!\!\frac{\proba{X = S_i}}{d\ t}  =  
\!\!\!\!\!\!\!\!\!\!\!\!\!\!\!\!\!\!\!\!\!\!\!\!\!\!\!\!\!
\!\!\!\!\sum_{ \ \ \ \ \ \ \ \ \ \ \ \ \ \ \ \ \ \ \longtransition{S_j}{S_i}{\alpha,r_{\langle\alpha,S_j,S_i\rangle}},\ S_j\in Y_{\mo{M}}}^{}
\!\!\!\!\!\!\!\!\!\!\!\!
r_{\langle\alpha,S_j,S_i\rangle}\proba{X=S_j}
\, 
\\[-2pt]
\ \ \ \ \ \ \ \ \ \ \ \ \ \ \ \ \ - \!\!\!\!\!\!\!
\!\!\!\!\!\!\!\!\!\!\!\!\!\!\!\!\!\!\!\!\!\!\!\!\!\!\!
\sum_{\ \ \ \ \ \ \ \ \ \ \ \ \ \ \longtransition{S_i}{S_j}{\alpha,r_{\langle\alpha,S_i,S_j\rangle}}\ ,\ S_j \in Y_\mo{M}}^{} \mkern-18mu \!\!\!\!\! 
\!\!\!\!\!\!\!\!\!\!\!\!\!\!\!\!
r_{\langle\alpha,S_i,S_j\rangle}\proba{X=S_i}
\end{multline}
The probability of being in a sub-chain $Y_i$ at time $t$ is equal to the sum of the probabilities of being in any of the states in $Y_i$:  $\proba{X \subset Y_i} = \sum_{S_i \in Y_i}^{}\ \proba{X=S_i}$.
To study these probability evolutions, we form one equation for each sub-chain $Y_i$. $\proba{X\subset Y_i}$ denotes the probability that at time $t$ the system is in any of the states in $Y_i$.
\begin{eqnarray}
\label{complexCKEquation}
\frac{d\  \proba{X\subset Y_i}}{d\ t} \!= \! \frac{d\ \sum_{S_i \in Y_i}^{}\!\proba{X=S_i} }{d\ t}\! =\!\!\! \sum_{S_i \in Y_i}^{}\!\!\!\frac{d\ \proba{X=S_i}}{d\ t}
\end{eqnarray}
In the context of $Y_i$, the probability fluxes which correspond to action types in $\actionsLargeModel{M}$ cancel each other and have no effect on the probability fluxes in and out of the whole sub-chain. 
Substituting (\ref{simpleCKEquation}) into (\ref{complexCKEquation}), using Lemma \ref{observation} and taking advantage of the assumption that the probability of being in boundary states is close to zero, we derive, $\forall\  Y_i \in Y_{\mo{M}}$: 
\begin{multline*}
\label{finalEquation}
\!\!\!\frac{d\ \proba{X \subset Y_i}}{d\ t}\!\approx\! -\!\!\!\! \sum_{\ Y_j\in Y_{\mo{M}}} \!\!\!\!\!r(Y_i,Y_j) \proba{X\subset Y_i}  
\\[-4pt]
+ 
\!\!\!\!\!\!\!
\sum_{\ \ Y_k \in Y_{\mo{M}}}
\!\!\!\!\!\! r(Y_k,Y_i) \proba{X\subset Y_k} 
\end{multline*}
which captures the probability evolutions at the level of sub-chains. The complete proof is described in \cite{technicalreport}.

For a large-scale PEPA model which satisfies Condition~\ref{aggregationCondition}, we have shown two approaches to deriving equations leading to the corresponding marginal probability distributions over sub-chains.
The two routes lead to the same equations since each sub-chain is represented as a single state in the aggregate CTMC and this CTMC is faithful in capturing the probability flows into and out of the sub-chains. However, it should be noted that the cost of deriving the aggregate ODEs and marginal distributions through the aggregated CTMC is orders of magnitude smaller than deriving them from the C-K equations, since this avoids the construction of the large system of C-K equations. 
In the next section, we consider an experiment where the accuracy of the approximate aggregate ODEs is investigated.

%% file: CheckingTheAccuracy/checkingTheAccuracy.tex
\section{Accuracy of the Aggregation}
\label{checkingTheAccuracy}

For a model which satisfies Condition \ref{aggregationCondition}, the accuracy of the marginal probability distributions derived via the aggregation, depends on the assumption that the probability mass in the  boundary states is close to zero. 
If, at all times, the small groups in the model are under heavy load and their cooperation capacity is saturated by the demand of the large groups, one would expect to get highly accurate approximate marginal probability distributions. 
Conversely, if the probability of being in boundary states, is not negligible, then the approximation method may lead to erroneous marginal distributions.  
In this section, we report on experiments to investigate the accuracy of the aggregation method in the context of the client-server model. 
\begin{figure}
\centering
\includegraphics[width=0.38\textwidth,height=0.19\textheight,clip=true,trim=155pt 270pt 125pt 280pt]{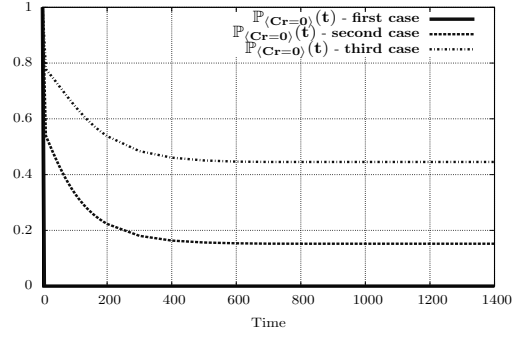}
\caption{Probability of being in boundary states ($\proba{C_r =0}$) for each case of the experiment.}
\label{threeCasesForCr}
\end{figure}

We consider Model~1 and assume that the small group, $\textit{Servers}$, contains 5 servers and the large group, $\textit{Clients}$, 100 clients. 
In the model's \emph{original} CTMC, which captures the dynamics of both clients and servers, the boundary states are those where the number of clients requesting service is zero (i.e.\ $C_r=0$).  
We constructed three versions of this model and across these, the following parameters were the same: $n_s\!=\!5$, $n_c\!=\!100$, $r_s\!=\!10$, $r_l\!=\!50$, $r_b\!=\!0.005$, $r_f\!=\!0.005$ and $r_c\!=\!\top$. 
However, in the first version $r_t=15$, in the second $r_t=0.2$ and in the third $r_t=0.1$.  
This change in $r_t$ causes a gradual increase in the probability of being in boundary states (see Fig.~\ref{threeCasesForCr}).
For each version, we calculated an approximate marginal probability distribution over the servers and compared it with a similar distribution derived from the exact analysis of the model's complete CTMC using the PRISM tool \cite{prism}.
There are multiple ways for comparing two probability distributions. For simplicity, we chose three different representative states from the distributions and compared the distributions only with respect to those particular states. 
Our comparison could readily be extended to the complete distributions. 
In this section, $Z$ denotes the stochastic process representing the client-server model's original CTMC and $Z_a$ denotes the stochastic process associated with the model's aggregated CTMC.



\begin{figure*}[t]
\centering
			\begin{subfigure}[b]{0.33\textwidth}
                \centering
                \includegraphics[width=\textwidth,height=0.2\textheight,clip=true,trim=155pt 270pt 125pt 280pt]{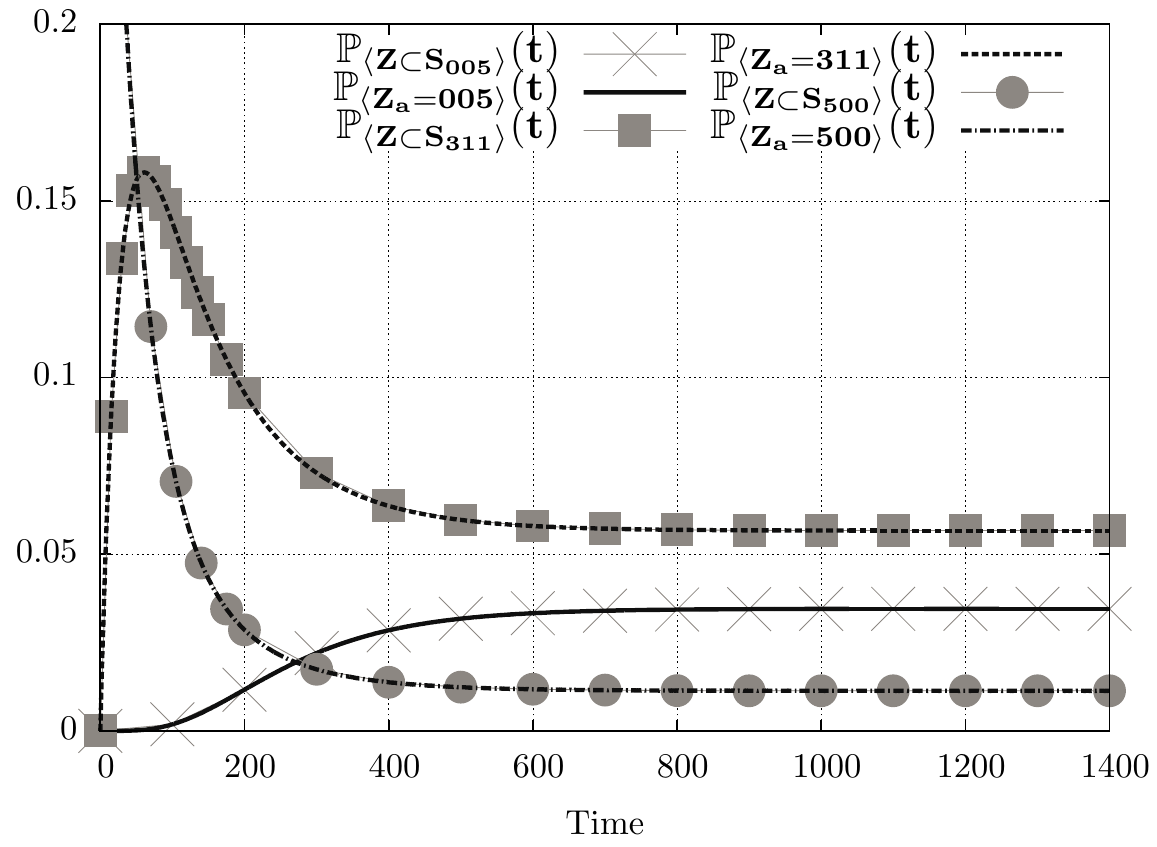}
                \caption{$\probNoT{Z\subset(5,0,0)},\probNoT{Z\subset(3,1,1},\probNoT{Z\subset(0,0,5)}$ first case.
}
                \label{fig:stateProbabilityFirstCase}
       \end{subfigure}
			\begin{subfigure}[b]{0.33\textwidth}
                \centering
                \includegraphics[width=\textwidth,height=0.2\textheight,clip=true,trim=155pt 270pt 125pt 280pt]{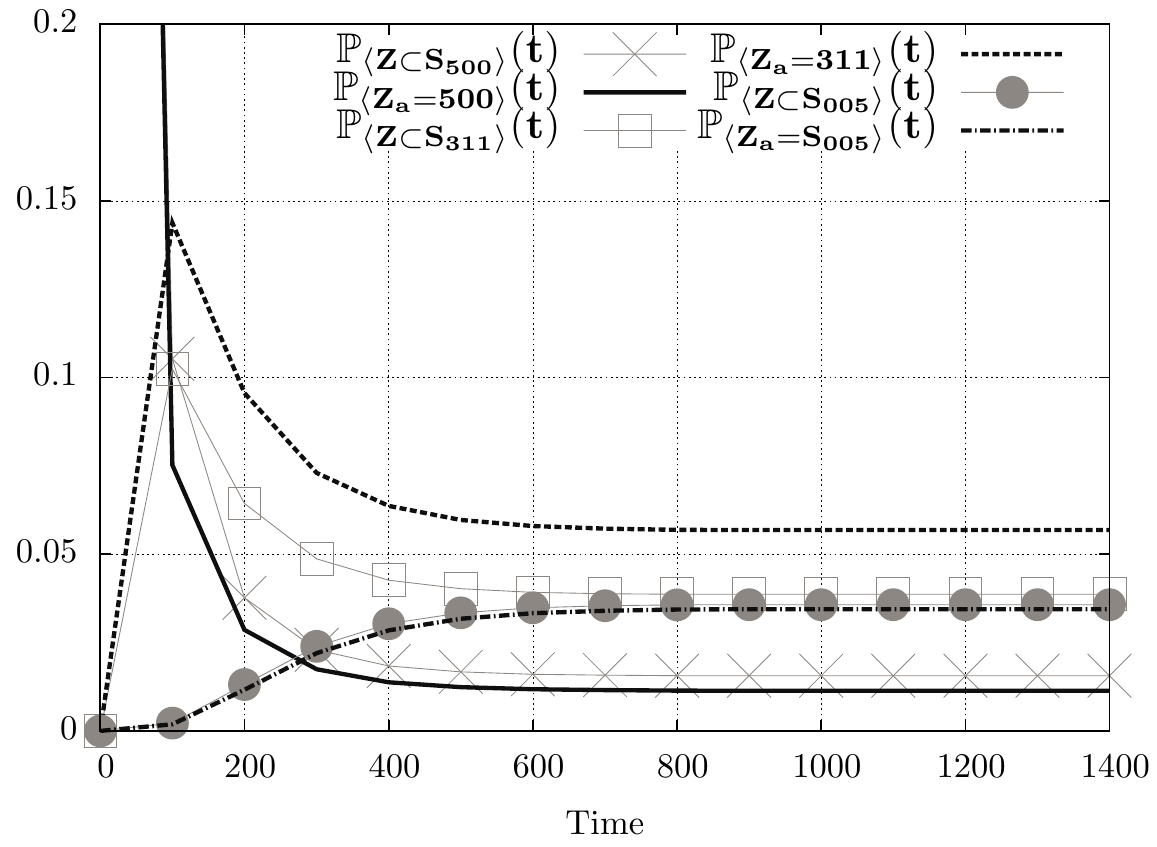}
                \caption{
                $\probNoT{Z\subset(5,0,0)},\probNoT{Z\subset(3,1,1)},\probNoT{Z\subset(0,0,5)}$ second case.
}
                \label{fig:stateProbabilitySecondCase}
       \end{subfigure}
       \begin{subfigure}[b]{0.32\textwidth}
                \centering
                \includegraphics[width=\textwidth,height=0.2\textheight,clip=true,trim=155pt 270pt 125pt 280pt]{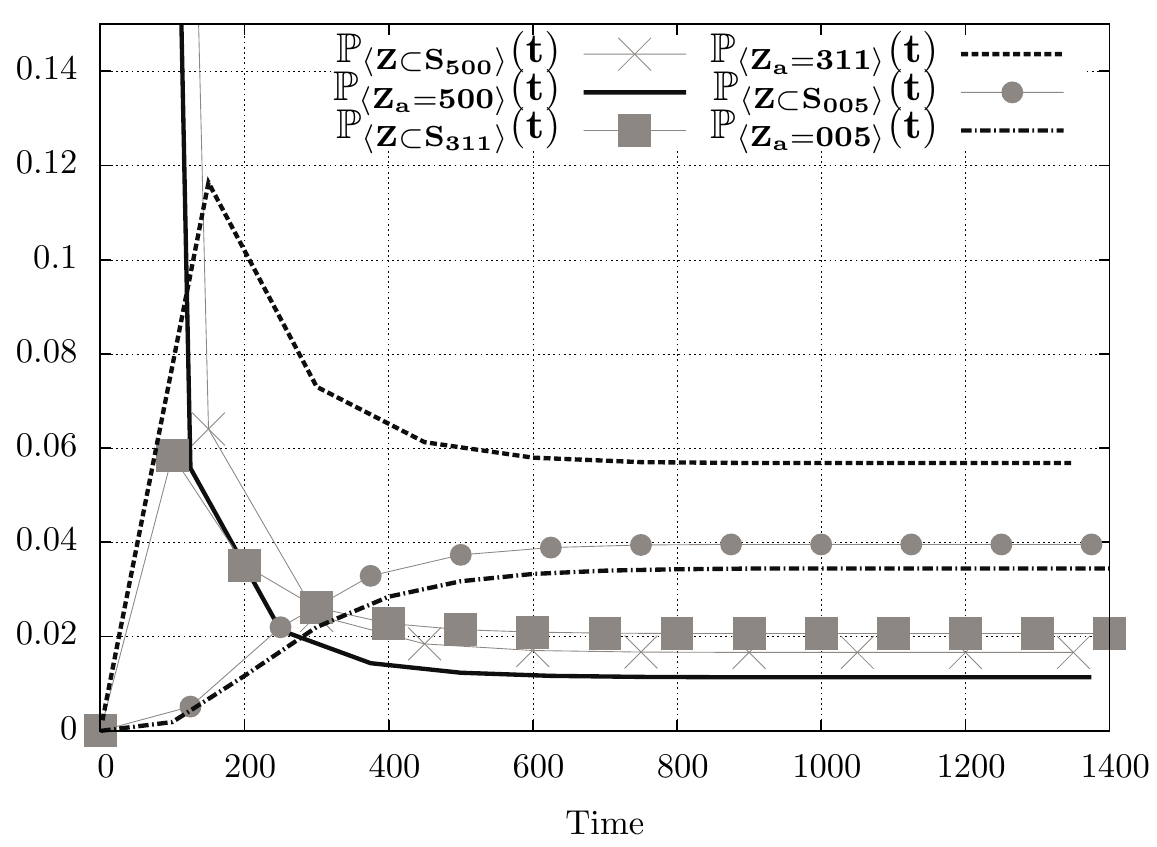}
                \caption{$\probNoT{Z\subset(5,0,0)},\probNoT{Z\subset(3,1,1)},\probNoT{Z\subset(0,0,5)}$ third case.
}
                \label{fig:stateProbabilityThirdCase}
       \end{subfigure}
%
			\caption{
			Comparison of exact and approximate probabilities of being in three representative states across the experiment cases.\label{fig:firstCase}
}
\end{figure*}

The parameters chosen for the first version ($r_t=15$) cause the servers to be under heavy contention at all times; i.e.\ probability of $C_r=0$ is close to zero.
Figure (\ref{fig:stateProbabilityFirstCase}) shows a comparison between probabilities calculated for three representative states, $\langle5,0,0\rangle$, $\langle3,1,1\rangle$ and $\langle0,0,5\rangle$, by the approximate and exact methods. 
As an example, $\proba{Z = (3,1,1)}$, denotes the probability that in the original CTMC, the system resides in a state where there are three idle, one logging and three broken servers and $\proba{Z_a = (3,1,1)}$, the probability that the aggregated CTMC resides in an equivalent state.

%

In the second case, $r_t=0.2$, thinking has a longer duration compared to the first case. 
This slows the flow of clients into the state of requesting communication with the servers. 
Thus, the probability of observing $C_r=0$ becomes higher as the servers' service capacity remains the same. 
The same measures were calculated for the second case and the results are reported in Fig.~(\ref{fig:stateProbabilitySecondCase}). 
Here, the probability of being in the boundary states is higher resulting in a less accurate the marginal distribution.
In the third case $r_t=0.1$, the probability of $C_r\!=\!0$ is relatively high (see Fig.~\ref{threeCasesForCr}). Hence, the deviation of the approximate distribution from the exact one is significantly larger than the previous cases. The outputs for this case are shown in Fig~(\ref{fig:stateProbabilityThirdCase}).

The aggregated CTMC can be used for deriving further performance indicators, such as dependability measures. 
Assume that the measure of interest is the number of working servers, i.e.\ those which are not broken.
Formally, let $E^{K}$ denote the probabilistic event that there are $K$ servers running. For instances, when K=5:
\begin{eqnarray*}
&&\!\!\!\!\!\!\!\!\!\!
\probNoT{Z \subset E^{5}}=\! \probNoT{Z\subset(5,0,0)} + \probNoT{Z\subset(4,1,0)} + \ldots\ +  \probNoT{Z\subset(0,5,0)} \approx\\
&&\!\!\!\!\!\!\!\!\!\!
\probNoT{Z_a \subset E^{5}}\!=\! \probNoT{Z_a=(5,0,0)} + \probNoT{Z_a=(4,1,0)} + \ldots\ +  \probNoT{Z_a=(0,5,0)}
\end{eqnarray*}

Using the aggregation method, the steady state values of $\probNoT{Z \subset E^{K}}, K=1,2,5$ and the previously presented outputs were also calculated and compared against the corresponding exact results (see Table~\ref{numericalPresentationOfTheError}). The comparison provides evidence that a higher probability of being in the model's boundary state corresponds to a less accurate aggregation. 
\newcommand{\probe}[1]{\mathbb{P}_{\langle #1 \rangle}}

\begin{table*}[t]
\begin{center}
\begin{tabular}{|c||c|c|c||c|c|c||c|c|c||} 
\hline 
 	&  \multicolumn{3}{|c||}{First Case} & \multicolumn{3}{|c||}{Second Case} & \multicolumn{3}{|c||}{Third Case} 
\\ \hline
	&	exact 	&	approximate	&	error (\%) 	 &	exact	& 	approximate	& 	error &		exact	& 	approximate 	& 	error(\%) 
\\ \hline
$\probe{Z_a=500} $	&	0.011 	&	0.011	&	0 	 &	0.015	& 	0.011	& 	26	 &	0.016	& 	0.011 	& 	31 
\\ \hline
$\probe{Z_a=311} $	&	0.056 	&	0.056	&	0 	 &	0.038	& 	0.056	& 	47 &		0.020	& 	0.056 	& 	180 
\\ \hline
$\probe{Z_a=005} $	&	0.034 	&	0.034	&	0 	 &	0.035	& 	0.034	& 	2 &		0.039	& 	0.034 	& 	12 
\\ \hline
$\probe{Z_a\subset E^{5}}$	&	0.028 	&	0.028	&	0 	 &	0.023	& 	0.028	& 	21 &	0.02	& 	0.028 	& 	40 
\\ \hline
$\probe{Z_a\subset E^{2}}$	&	0.310 	&	0.317	&	2 	 &	0.328	& 	0.317	& 	3	 &		0.336	& 	0.317 	& 	5.6 
\\ \hline
$\probe{Z_a\subset E^{1}}$	&	0.161 	&	0.165	&	2.48 	 &	0.171	& 	0.165	& 	3.5	 &		0.189	& 	0.165 	& 	12 
\\ \hline
\end{tabular}
\caption{Comparion of the equilibrium probabilities \label{numericalPresentationOfTheError} calculated separately by the approximate and exact methods.}
\end{center}
\end{table*}

In this experiment, using a quad-core machine with 2G of RAM, for each case, deriving the approximate steady state marginal distribution through the aggregated CTMC took 10--15 seconds. Deriving the same distribution by the exact analysis took nearly 650--700 seconds. The former takes advantage of the possibility of aggregation whereas the latter, derives the marginal distribution from a state space where the detailed dynamics of the clients are also captured.

Applying the aggregation method relies on prior knowledge about the model's behaviour with respect to its boundary states. 
Such knowledge can be supplied by the domain experts or by monitoring the real system. 
In large-scale resource-bound communication networks, 
the resources are almost continuously under contention; e.g.\ processing input transactions or dealing with a frequent incoming flow of packets. 
Thus, for such systems, the probability of resources waiting for the users is close to zero and the aggregation algorithm proposed can be safely used.

%% file: RelatedWork/new-relatedWork.tex
\section{Related Work} 
\label{relatedWork}

Many aggregation and decomposition techniques have been developed to address the problem of analysing large-scale models. 
These methods usually exploit the structure of the model's CTMC and allow for a timely analysis which otherwise would have been impossible. 
Our aggregation method can be viewed as most closely related to two existing concepts: \emph{quasi-separability} and \emph{lumpability}.

The method of quasi-separability was originally developed for queueing networks \cite{Mitrani1994151} and later extended to Markovian process algebra models \cite{Thomas1998}. 
Given a process, the modeller categorises each component as being either an \emph{environment} component or an \emph{internal} one. 
If the model satisfies a \emph{quasi-separability condition}, a number of sub-models are generated where each sub-model captures the dynamics of a number of internal components in combination with all environment ones. 
For internal components, their configuration is probabilistically independent of the other internal ones and therefore, their behaviour can be studied \emph{in separation} via only their own sub-model. 
Using this method for a given model, 
for each sub-model one marginal distribution is derived. 

We investigated the quasi-separability condition in the context of PEPA population models. 
In particular, for a model in which there is an interaction between small and large groups, we established the conditions which determine whether the small groups evolve independently from the large ones and if their dynamics be studied \mbox{in isolation}. 

The second concept is lumpability (ordinary lumpability), which forms a partition over the model's CTMC satisfying a condition on rates. 
Using lumpability, the model's aggregated CTMC has a single state for each partition. 
The size of the aggregated CTMC is smaller than the original one and therefore, it provides the means for efficient performance analysis. 
Our method follows a similar approach. 
However, it is important to note that our conditions are less stringent than those for lumpability, and will aggregate a larger class of models. 
More specifically, the presence of boundary states in the CTMC of a model may break the lumpability condition whereas our method can handle such a CTMC\@.

%% file: Conclusion/new-conclusion.tex
\section{Conclusion}
\label{conclusion}
In this paper we considered a sub-class of large-scale PEPA models. 
These models consist of groups with non-uniform population sizes; one or more small groups cooperating with some large ones. 
These models reflect the situation in many of existing systems where a small population of resources are used by a significantly larger population of users.
We showed that if the model satisfies two conditions, an approximate aggregated CTMC can be constructed and used to efficiently derive certain performance metrics. 
These are the conditions. First, the rate of all cooperation between the model's small groups and large ones should be controlled solely by the small groups. 
Second, as the model executes, it must be unlikely to experience its boundary states, i.e.\ the ones where the small groups are blocked while waiting for the cooperation with the large groups. 
We proposed an algorithm which, for a conforming model, directly derives its aggregated CTMC without constructing the original CTMC\@. 
The aggregated CTMC captures, in a detailed manner, the dynamics of the model's small groups and can be used to derive an approximate marginal probability distribution over those groups. 
Calculating such a distribution using the aggregated CTMC is orders of magnitude faster than deriving the same distribution from the analysis of the original CTMC. 


\textbf{Future Work}. For a conforming model, we showed how its aggregated CTMC can be used to study the behaviour of its small groups. 
However, the aggregated CTMC, as a compact representative of the model's original CTMC, can be exploited more widely. 
The original CTMC and the aggregated CTMC can be shown to share certain characteristics, yet investigating them in the original CTMC may be prohibitively expensive. 
Therefore, our aim is to investigate how to establish important characteristics, such as near complete decomposability in the aggregated CTMC and then exploit them in the original CTMC\@.